\documentclass{aa}  
\usepackage{graphicx}
\usepackage{subfigure}
\usepackage{txfonts}
\usepackage{natbib}
\bibpunct{(}{)}{;}{a}{}{,} 
\begin{document}
\title{Pulsating low-mass white dwarfs in the frame of new evolutionary 
sequences} 
\subtitle{VI. Thin H-envelope sequences and
  asteroseismology of ELMV stars revisited}
\author{Leila M. Calcaferro\inst{1,2},
        Alejandro H. C\'orsico\inst{1,2},  
        Leandro G. Althaus\inst{1,2},  
        Alejandra D. Romero\inst{3}, \and
        S. O. Kepler\inst{3}}
\institute{$^{1}$ Grupo de Evoluci\'on Estelar y Pulsaciones,
  Facultad de Ciencias Astron\'omicas y Geof\'isicas,
  Universidad Nacional de La Plata, Paseo del Bosque s/n,	
  1900, La Plata, Argentina\\
  $^{2}$ Instituto de Astrof\'isica La Plata,
  CONICET-UNLP, Paseo del Bosque s/n, 1900, La Plata, Argentina\\
  $^{3}$ Instituto de Física, Universidade Federal do Rio Grande do Sul, 
  Av. Bento Goncalves 9500, Porto Alegre 91501-970, RS, Brazil\\
           \email{lcalcaferro,acorsico,althaus@fcaglp.unlp.edu.ar; alejandra.romero@ufrgs.br,kepler@if.ufrgs.br}     
           }

\date{Received ; accepted }

\abstract {Some low-mass white-dwarf (WD) stars with H atmospheres,
  which are currently being detected in our galaxy, show long-period
  $g$(gravity)-mode pulsations, and comprise the class of
  pulsating WDs called extremely low-mass variable (ELMV) stars.
  At present, it is generally
  believed that these stars have thick H envelopes. However, from stellar
  evolution considerations, the existence of low-mass WDs with
  thin H envelopes
  is also possible.}{We present a thorough asteroseismological analysis of
  ELMV  stars on the basis of a complete set of fully evolutionary
  models that represents low-mass He-core WD stars harboring
  a range of H envelope thicknesses. Although there are currently nine ELMVs,
    here we
  only focus on those that exhibit more than three periods and whose periods do not
  show significant uncertainties.}{We considered $g$-mode adiabatic 
  pulsation periods for low-mass He-core WD models with stellar
  masses in the range $[0.1554-0.4352]\ M_{\sun}$, effective
  temperatures in the range $[6000-10000]\ $K, and H envelope
  thicknesses in the interval $-5.8 \lesssim \log(M_{\rm H}/M_{\star})
  \lesssim -1.7$. We explore the effects of employing different 
  H-envelope thicknesses on
  the adiabatic pulsation properties of  low-mass He-core WD
  models, and perform period-to-period fits to ELMV stars
  to search for a representative asteroseismological model.}{We found
  that the mode-trapping effects of $g$ modes depend sensitively on the
  value of $M_{\rm H}$, with the trapping cycle and trapping amplitude
  larger for thinner H envelopes. We also found that the asymptotic period
  spacing, $\Delta \Pi^{\rm a}$, is longer for thinner H envelopes. Finally,  
  we found asteroseismological models (when possible) for the stars under
  analysis, characterized by canonical (thick) and by thin H envelope.
  The effective temperature and stellar mass of these models are in agreement
  with the spectroscopic 
  determinations.}{The fact that we have found asteroseismological solutions
  with H envelopes thinner than canonical gives a clue of the possible
  scenario
  of formation of these stars. Indeed, in the light of our results, some of
  these stars could have been formed by binary evolution through unstable mass
  loss.}  
  \keywords{asteroseismology --- stars:
  oscillations ---  white dwarfs --- stars: evolution --- stars:
  interiors}
  \authorrunning{Calcaferro et al.}
  \titlerunning{Asteroseismology of ELMV stars}
  \maketitle
%

\section{Introduction}
\label{introduction}

The vast majority of stars,  including our Sun,
will end up their lives as WD
stars \citep{2008ARA&A..46..157W,
  2008PASP..120.1043F,2010A&ARv..18..471A}. Most WDs
\citep[$\sim 85\%$; see][]{2016MNRAS.455.3413K} show hydrogen (H) in their
atmospheres, and are classified
spectroscopically as DA WDs. The average mass of the DA WDs is
$\sim 0.64 M_{\sun}$ \citep{2017EPJWC.15201011K} and they probably harbor
carbon-oxygen (CO) cores. There are also very massive WDs
($M_{\star}\gtrsim  1.05 M_{\sun}$) with oxygen-neon (ONe) cores,
and, at the other extreme of mass range, WDs with low mass
($M_{\star} \lesssim 0.45 M_{\sun}$), which are believed to have cores made
of helium (He). Present day He-core WDs are supposed to be the outcome of
strong mass-loss episodes in interactive binary systems,
before the ocurrence of the He flash at the red giant branch (RGB)
phase of low-mass
stars \citep[see][for instance]{2013A&A...557A..19A,
  2016A&A...595A..35I}. Nowadays, this
evolutionary scenario is the most likely mechanism for the formation of
the extremely low-mass (ELM) WDs, with masses below
$\sim 0.18- 0.20 M_{\sun}$. At present, there is no agreement among
researchers as for the precise upper-mass limit for ELM WDs.
The value we propose here and in our previous works
($M_{\star} \lesssim 0.18-0.20 M_{\sun}$) is a physically-motivated limit,
because it refers to WDs that (i) have not experienced CNO flashes
in their past evolution, (ii) are characterized by very long cooling timescales,
and (iii) have pulsational properties quite different as compared with
the systems
that experienced flashes \citep[see][]{2013A&A...557A..19A,2014A&A...569A.106C}.
Nevertheless, this limit depends on the WD progenitors metallicity
\citep[][]{2016A&A...595A..35I}. Other authors prefer to adopt a
value of $\sim 0.3 M_{\sun}$ as the
upper mass limit for ELM WDs \citep[e.g.,][]{2016ApJ...818..155B}.

    In the last decade, several low-mass WDs, including ELM
WDs, have been discovered with the ELM, SPY
and WASP surveys \citep[see][for instance]{2009A&A...505..441K,  2010ApJ...723.1072B,
  2012ApJ...744..142B, 2011ApJ...727....3K,
  2012ApJ...751..141K,
  2015MNRAS.446L..26K, 2015ApJ...812..167G, 2016ApJ...818..155B,
  2017ApJ...847...10B}, and some of them
have been found to exhibit multi-periodic
brightness variations compatible with $g$(gravity)-mode pulsations
\citep{2012ApJ...750L..28H,
  2013ApJ...765..102H,2013MNRAS.436.3573H,
  2015MNRAS.446L..26K,2015ASPC..493..217B,2017ApJ...835..180B,
  2018MNRAS.478..867P,2018arXiv180511129B}.
These pulsating low-mass WDs constitute the new class of
variable WDs, generically named ELMV stars. ELMVs provide us an
unique chance to dig into the interiors of these
stars, and ultimately to test the scenarios of their formation
by employing WD asteroseismology
\citep[][]{2008ARA&A..46..157W,2008PASP..120.1043F,2010A&ARv..18..471A}. In particular,
\cite{2010ApJ...718..441S,2012A&A...547A..96C,2014A&A...569A.106C}
demonstrated that $g$ modes in ELMVs are  mainly confined to the
core regions, at variance with the case of average-mass
CO-core pulsating DA WDs (DAV or ZZ Ceti stars). This, in principle,
could allow one to put constraints to the core chemical structure
of ELMV stars. 

Asteroseismology applied to WDs has already proven to be successful
for peering into the interior of these stars \citep{2008ARA&A..46..157W,
  2008PASP..120.1043F,2010A&ARv..18..471A}. In particular, two
main asteroseismological avenues have been employed: one considering
stellar models harboring parametrized chemical composition profiles, and
another involving fully evolutionary models characterized
by chemical profiles resulting from all the processess experienced during
the evolution of the WD progenitors. The former approach constitutes
a powerful forward method with the flexibility of allowing a
full exploration of the parameter space
(the total mass, the mass of the H and He envelopes, the thickness
of the chemical transition regions, the core 
chemical structure and composition, etc) to find an optimum
asteroseismological model \citep[see][among others]{1998ApJS..116..307B,
  2001ApJ...552..326B, 2006A&A...446..223P,2006A&A...453..219P,
  2013MNRAS.432..598P,2016MNRAS.461.4059B, 2016ApJS..223...10G,
  2017A&A...598A.109G,2017ApJ...834..136G,2018Natur.554...73G}. 
  The weak point of this
  method is that in view of the lack of a large number of observed
  pulsations it can lead to asteroseismological solutions
characterized by chemical structures that are not predicted by any
scenario of WD evolution. Just to give an example,
for ZZ Ceti stars, the derived asteroseismological
models may have a pure C buffer, which is difficult to predict by the currently accepted 
channels of WD formation, or not realistic abundances of C and O at
the core that are at variance with the current uncertainty of the 
$^{12}$C$(\alpha,\gamma)^{16}$O reaction rate. The second avenue 
was developed at La Plata Observatory and
utilizes the fully evolutionary models that result from the
complete evolution of the progenitor stars, starting at the Zero Age Main
Sequence (ZAMS) all the way down to the WD phase.
This method has been applied to GW Virginis (pulsating PG1159) stars
\citep{2007A&A...461.1095C,2007A&A...475..619C,
  2008A&A...478..869C,2009A&A...499..257C,2014MNRAS.442.2278K,
  2016A&A...589A..40C}, and also to DBV WDs
(He-rich atmosphere) \citep{2012A&A...541A..42C,2014A&A...570A.116B,
  2014JCAP...08..054C}.
Regarding ZZ Ceti stars, this avenue has been successfully employed
by \cite{2012ApJ...757..177K,2012MNRAS.420.1462R,2013ApJ...779...58R,
2017ApJ...851...60R}. In this approach, the chemical
structure of the equilibrium models is consistent with the pre-WD
evolution. However, there are important uncertainties related to
 the evolutionary
processes that take place during the evolution of the progenitor star,
like the exact amount of overshooting, the precise number of
thermal pulses during the TP-AGB phase, the value of the
$^{12}$C$(\alpha,\gamma)^{16}$O nuclear
reaction rate which is relevant during the central He burning stage, mass
loss rates, etc. Recently, in order to determine the influence of these
uncertainties on the properties of asteroseismological models of ZZ
Ceti stars derived with this method, an assessment has been carried out 
\citep{2015ASPC..493..225D, 2017A&A...599A..21D, 2018A&A...613A..46D},
and the results  indicate that the impact on asteroseismological models
is well quantifiable and bounded. 

In \cite{2017A&A...607A..33C}, we have performed an asteroseismological
analysis for the first time to all the known ELMVs,
whose spectroscopic parameters and pulsation periods are listed
in Tables 1 to 9 of that paper. In that paper, 
period-to-period fits to the target stars were carried out employing
adiabatic radial  and non-radial $g-$ and $p-$mode pulsation periods
of low-mass He-core WD evolutionary models with stellar masses
between $0.1554$ and $0.4352\ M_{\sun}$, resulting from the computations
of \cite{2013A&A...557A..19A}, that take into account the binary
evolution of the progenitor  stars. Although the stars under study
show few resolved periods and there are multiple possible solutions
to the period fits, \cite{2017A&A...607A..33C}
found that for most cases a seismological model can be adopted, and
the corresponding values of $M_{\star}$ and $T_{\rm eff}$ lie within
the expected
spectroscopic determinations. Also, they found that in general,
the pulsation periods corresponding to the asteroseismological
models are associated with pulsationally unstable eigenmodes,
according to their nonadiabatic
computations. However, these authors noted that for most of the
stars analyzed, the derived asteroseismological models are more
massive in comparison with the spectroscopic results. The authors
concluded that this tendency could be related to some extent to the fact
that they only considered low-mass He-core WD models characterized
by outer H envelopes coming from the
stable mass loss scenario via Roche-lobe overflow,
instead of considering the possibility that they may have thinner H
envelopes
\citep[see the recent works by][]{2013A&A...557A..19A,
  2016A&A...595A..35I}.
In this connection, it cannot be discarded the existence of such 
low-mass WDs that are unable to sustain residual H burning, and whose
formation may be the result of common-envelope evolution of close binary systems 
\citep{2016MNRAS.460.3992N,2016MNRAS.462..362I,2017MNRAS.470.1788C},
or from the lost of the envelope of a RGB
star induced by an inspiralling giant planet
\citep{1998A&A...335L..85N,2002PASP..114..602D,2017arXiv170608897S}.
In view of these considerations, we conclude that the existence of ELM WDs with thin H envelopes must be considered.

The present work is the sixth part of a series
\citep{2014A&A...569A.106C,2016A&A...585A...1C,
  2016A&A...588A..74C,2017A&A...600A..73C} devoted to the pulsational
properties of low-mass WD and pre-WD stars. Throughout them, studies of
the adiabatic properties and nonadiabatic pulsation stability analyzes
of ELMV and pre-ELMV pulsating stars have been performed. Also, 
the theoretical rates of period change of ELMV and pre-ELMV stars has
been assessed. In the fifth part \citep{2017A&A...607A..33C},
we performed a detailed asteroseismological study of the complete set of
confirmed ---and alleged--- ELMV stars. In this work, we repeat the
analysis carried out in \cite{2017A&A...607A..33C},
but this time incorporating new evolutionary sequences of low-mass He-core WDs
including thin H envelope models. This allows us to expand the
parameter space of our asteroseismological analysis by adopting also
the thickness
of the H envelope ($M_{\rm H}$) as a free parameter, in addition
to $M_{\star}$ and $T_{\rm eff}$. Specifically, we perform asteroseismological
period fits to ELMV stars,
employing adiabatic pulsation periods of $g$ modes corresponding to a
big set of He-core WD models with stellar masses in
the range $0.1554 \lesssim M_{\star}/M_{\sun} \lesssim 0.4352$,
effective temperatures in the range $6000 \lesssim T_{\rm eff}
\lesssim 10000$ K, and H envelope thicknesses in
the interval $-5.8 \lesssim \log(M_{\rm H}/M_{\star}) \lesssim -1.7$.

The paper is organized in the following way. A brief summary of the
numerical codes and the stellar models employed is provided in
Sect. \ref{numerical} and \ref{models}.
In Sect. \ref{single-double}, we show 
that for ELM WD models with
thin H envelopes, single- and double-layered chemical structures
for the H envelope are expected, depending on the value of the
envelope thickness. In Sect. \ref{pulsation-properties}
we describe the impact that the employment of thin H
envelopes has on the adiabatic pulsation properties of ELMV WD models.
Next, in Sect. \ref{astero}, we search for
the best-fit asteroseismological model by comparing the individual
periods from each ELMV star under analysis with theoretical periods
from our grid of
models.  Finally, in Sect. \ref{conclusions} we summarize the main
findings of this work. 

\section{Numerical codes}
\label{numerical}

\begin{table*}
\centering
\caption{Stellar masses of our set of low-mass He-core WD models (column 1) and the H content associated with different thicknesses of the envelope adopted for each stellar mass ($T_{\rm eff} \sim 8000$ K).
Column 2 indicates the upper limit of the H-envelope thickness (``canonical envelope'')
for each stellar mass as given by our fully evolutionary computations that assume that
mass loss proceeds via stable Roche lobe
overflow.}
\begin{tabular}{l|ccccccc}
\hline
\hline
$M_{\star}/M_{\odot}$ & $\log(M_{\rm H}/M_{\star})$  & $\log(M_{\rm H}/M_{\star})$ &$\log(M_{\rm H}/M_{\star})$ &$\log(M_{\rm H}/M_{\star})$ &$\log(M_{\rm H}/M_{\star})$ &$\log(M_{\rm H}/M_{\star})$ &$\log(M_{\rm H}/M_{\star})$ \\
\hline
$0.1554$ &  $-1.69$ & $-2.50$ & $-3.00$ & $-3.72$ & $-4.37$ & $-5.02$ & $-5.20$ \\
$0.1612$ &  $-1.76$ & $-2.49$ & $-3.00$ & $-3.70$ & $-4.35$ & $-5.04$ & $-5.23$ \\
$0.1650$ &  $-1.82$ & $-2.50$ & $-3.01$ & $-3.69$ & $-4.35$ & $-5.06$ & $-5.28$ \\
$0.1706$ &  $-1.89$ & $-2.50$ & $-3.01$ & $-3.69$ & $-4.39$ & $-5.13$ & $-5.32$ \\
$0.1762$ &  $-1.95$ & $-2.50$ & $-3.00$ & $-3.66$ & $-4.34$ & $-5.14$ & $-5.43$ \\ 
$0.1805$ &  $-2.44$ &  $-$    & $-3.00$ & $-3.65$ & $-4.33$ & $-5.17$ & $-5.51$ \\
$0.1869$ &  $-2.37$ &  $-$    & $-2.99$ & $-3.64$ & $-4.32$ & $-5.19$ & $-5.61$ \\
$0.1921$ &  $-2.35$ &  $-$    & $-3.02$ & $-3.62$ & $-4.30$ & $-5.17$ & $-5.58$ \\
$0.2025$ &  $-2.43$ &  $-$    & $-3.00$ & $-3.61$ & $-4.30$ & $-5.20$ & $-5.62$ \\
$0.2390$ &  $-2.45$ &  $-$    & $-3.04$ & $-3.67$ & $-4.28$ & $-5.15$ & $-5.58$ \\
$0.2707$ &  $-2.96$ &  $-$    &   $-$   & $-3.67$ & $-4.32$ & $-5.19$ & $-5.54$ \\
$0.3205$ &  $-2.81$ &  $-$    &   $-$   & $-3.60$ & $-4.30$ & $-5.08$ & $-5.54$ \\
$0.3624$ &  $-3.10$ &  $-$    &   $-$   & $-3.62$ & $-4.32$ & $-5.15$ & $-5.64$ \\
$0.4352$ &  $-3.21$ &  $-$    &   $-$   & $-3.71$ & $-4.32$ & $-5.14$ & $-5.79$ \\
\hline 
\end{tabular}
\label{table_01}
\end{table*}

As in the previous papers of this series, in the present work we have
made use of the evolutionary models  of
low-mass He-core WDs coming from progenitors with metallicity of
$Z = 0.01$,
generated with  the {\tt LPCODE} stellar
evolution code, following the procedure described thoroughly  in
\cite{2013A&A...557A..19A}.  {\tt LPCODE} evolutionary code computes
the complete evolutionary stages which lead to the WD formation. In this way,
it allows the study of the evolution of the WD consistently
with the predictions of the progenitors evolutionary history.
A complete description of the input physics
of {\tt LPCODE} is given in \cite{2013A&A...557A..19A} and
references therein. We refer the interested reader to that paper for details.
It is worth mentioning that time-dependent diffusion due to gravitational settling
and chemical and thermal diffusion of nuclear species  was considered, following  the
multicomponent  gas  treatment  of \citet{1969fecg.book.....B}.

Adiabatic pulsation periods for non-radial dipole ($\ell= 1$) and
quadrupole ($\ell= 2$)
$g$ modes were taken from \cite{2014A&A...569A.106C}
in the case of WD models with canonical H envelope thicknesses. For WD models
with thinner H envelopes, the periods were computed specifically
for the present work. In both cases, the pulsation periods were computed
employing the adiabatic version of the {\tt LP-PUL} pulsation code
\citep{2006A&A...454..863C}. The Brunt-V\"ais\"al\"a
frequency ($N$) was computed following the Ledoux Modified
treatment \citep{1990ApJS...72..335T,1991ApJ...367..601B}.

\section{Model sequences}
\label{models}

In our analysis, we employed realistic configurations for
low-mass He-core WD stars computed by \cite{2013A&A...557A..19A}
by imitating the binary evolution of the progenitor stars assuming
initial configurations consisting of a $1.0 M_{\sun}$ Main Sequence (donor)
star and a $1.4 M_{\sun}$ neutron star
companion as the other component. By varying the initial orbital periods
at the beginning of the Roche lobe phase between $0.9$ and $300\ $d, 
a total of 14 initial He-core WD
models are obtained, with the following stellar masses: $0.1554$, $0.1612$,
$0.1650$, $0.1706$,
$0.1762$, $0.1805$, $0.1863$, $0.1921$, $0.2025$,  $0.2390$,
$0.2707$,  $0.3205$, $0.3624$ and $0.4352\ M_{\sun}$.
The evolution of these models was
computed down to the range of luminosities of cool WDs, including the
stages of multiple thermonuclear CNO flashes during the beginning of
the cooling branch (see \cite{2013A&A...557A..19A} for details about
the procedure adopted to carry this computations on).
We stress that mass loss proceeds here via stable Roche lobe overflow.
Hence, WD remnants with \emph{thick} H envelopes are expected
\citep[see][]{2013A&A...557A..19A}.

\begin{figure} 
\begin{center}
\includegraphics[clip,width=9 cm]{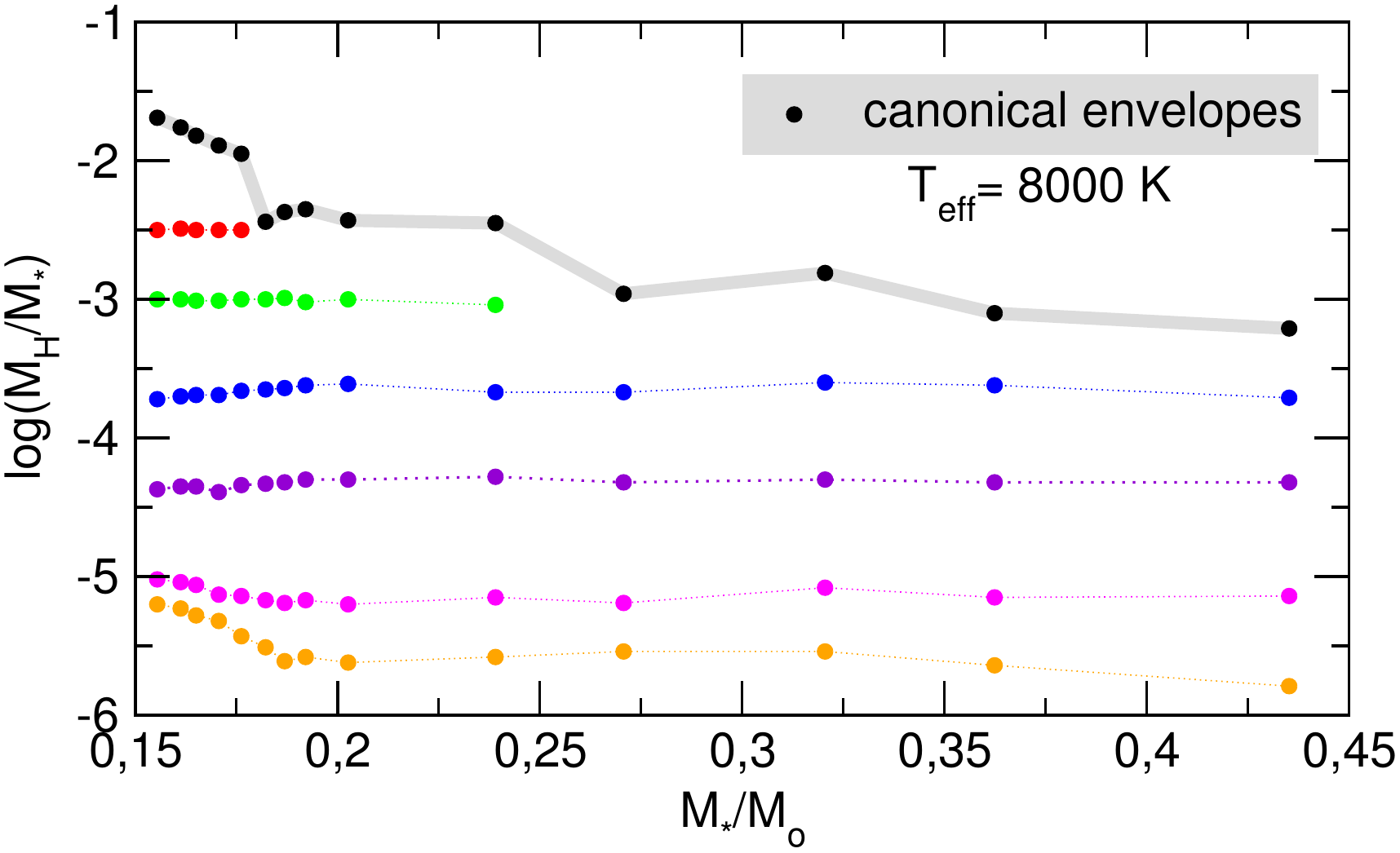} 
\caption{Grid of low-mass He-core WD evolutionary sequences considered in
  this work shown in the $M_{\star}-\log(M_{\rm H}/M_{\star})$ plane.
  The small circles represent each sequence of WD models with a given
  stellar mass and a specific thickness of the H envelope
  at $T_{\rm eff} \sim 8000$ K. The black circles connected
  with a thick (gray) line  correspond  to  the  values  of  the  maximum
  H  envelope thickness as predicted by the evolutionary
  computations of \cite{2013A&A...557A..19A}. For each sequence,
  we have pulsationally analyzed
  about 200 stellar models covering the interval of $T_{\rm eff}$
  between $6000$ and $10000$ K.}
\label{figure_01} 
\end{center}
\end{figure}

A novel aspect of this paper is the inclusion of new evolutionary sequences
of low-mass He-core WDs with thin H envelopes. 
The consequences of the presence of thin H envelopes on the cooling
ages of low-mass He-core WDs have been studied recently in
 \cite{2018A&A...614A..49C}. The procedure we follow to produce these
 new model sequences is straightforward \citep[see the case of ZZ
   Ceti stars in][]{2012MNRAS.420.1462R}.
For the purpose of getting a range of H envelope thicknesses, for  each
sequence  characterized  by  a given value of  $M_{\star}$  and  a
thick  (canonical) value  of $M_{\rm H}$,  as  predicted  by  the
 computation  of  the  pre-WD evolution  (second  column
of  Table  \ref{table_01}), we made the replacement of
$^1$H by $^4$He from a given mesh point, to obtain certain desired values 
of the H envelope mass. This artificial procedure is done at very
high $T_{\rm eff}$ values at the final cooling track 
to wash out any unphysical transitory effects associated to this
procedure long before the models reach
the pulsating stage of ELMV WD stars. After changing the thickness of the H
envelope, we allowed time-dependent element diffusion to act
while the WD models cool down until they get to the typical values of
$T_{\rm eff}$ that represent the ELMV instability
strip ($T_{\rm eff} \sim 10000\ $K). Diffusion strongly erodes the chemical
profiles at the He/H chemical transition regions. The values of the H
content that result for the different envelope thicknesses of
WD models at $T_{\rm eff}\sim 8000\ $K are displayed in
Table \ref{table_01}. Also, a  graphical
representation of the grid of models employed in this paper
is displayed in Fig. \ref{figure_01}, where a thick (gray) line
connects the canonical values of $M_{\rm H}$ as predicted by stellar evolution.
Our augmented grid of models has a total of 85
sequences of low-mass He-core WD models, which contains
$\sim 17000$ stellar models that were pulsationally analyzed.

\begin{figure} 
\begin{center}
\includegraphics[clip,width=9 cm]{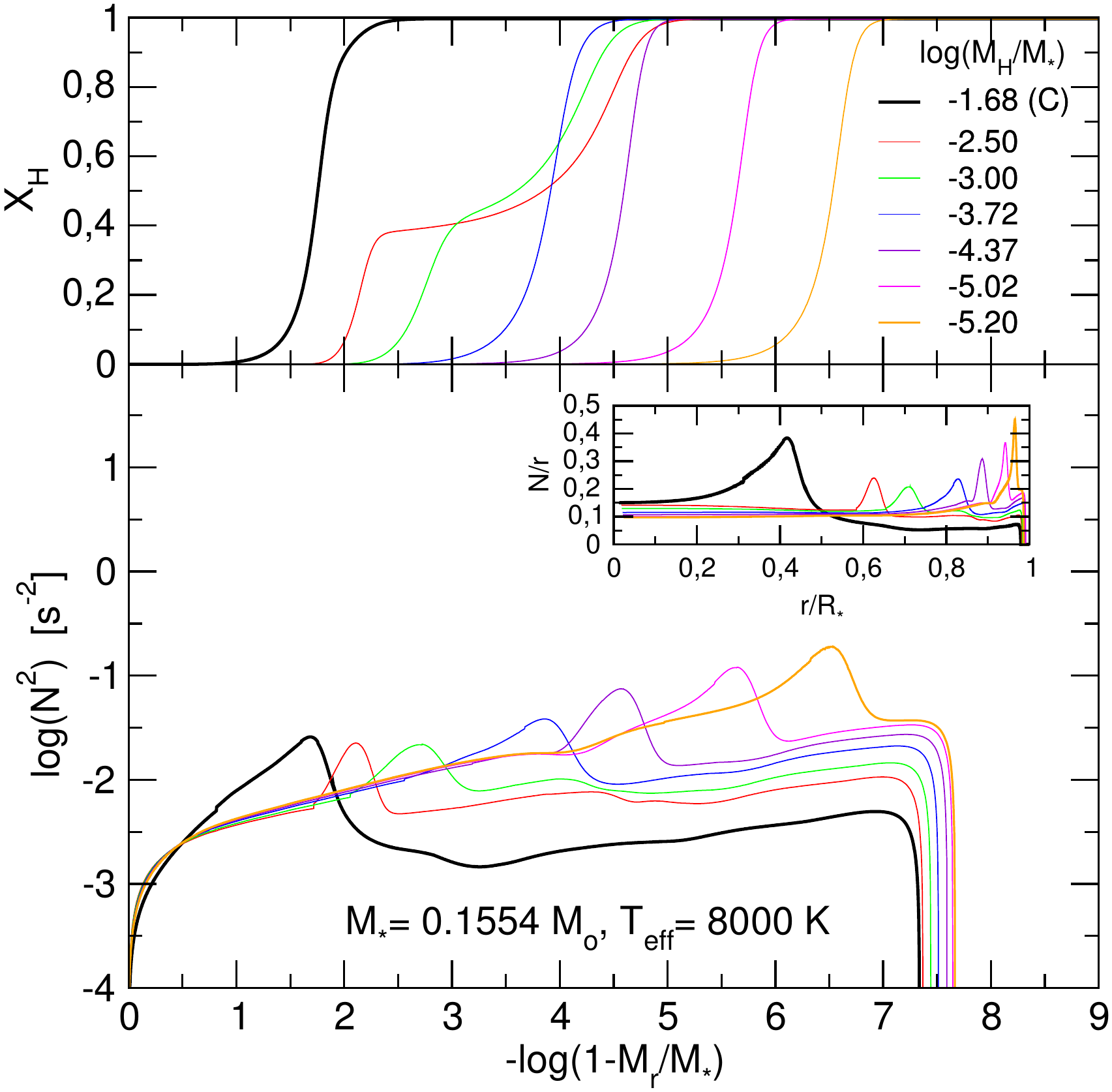} 
\caption{Chemical profiles of H for WD models with
  $M_{\star}= 0.1554 M_{\sun}$, $T_{\rm eff} \sim 8000$ K and
  several thicknesses of H envelope (upper panel). Thick black line
  corresponds to the canonical envelope.  Run of the
  logarithm of the squared Brunt–V\"ais\"al\"a frequency for each
  depicted model (lower panel). 
  The inset shows the quantity $N/r$ as a function of
  the radial coordinate, $r$, for the same WD models.} 
\label{figure_02} 
\end{center}
\end{figure}

\begin{figure} 
\begin{center}
\includegraphics[clip,width=9 cm]{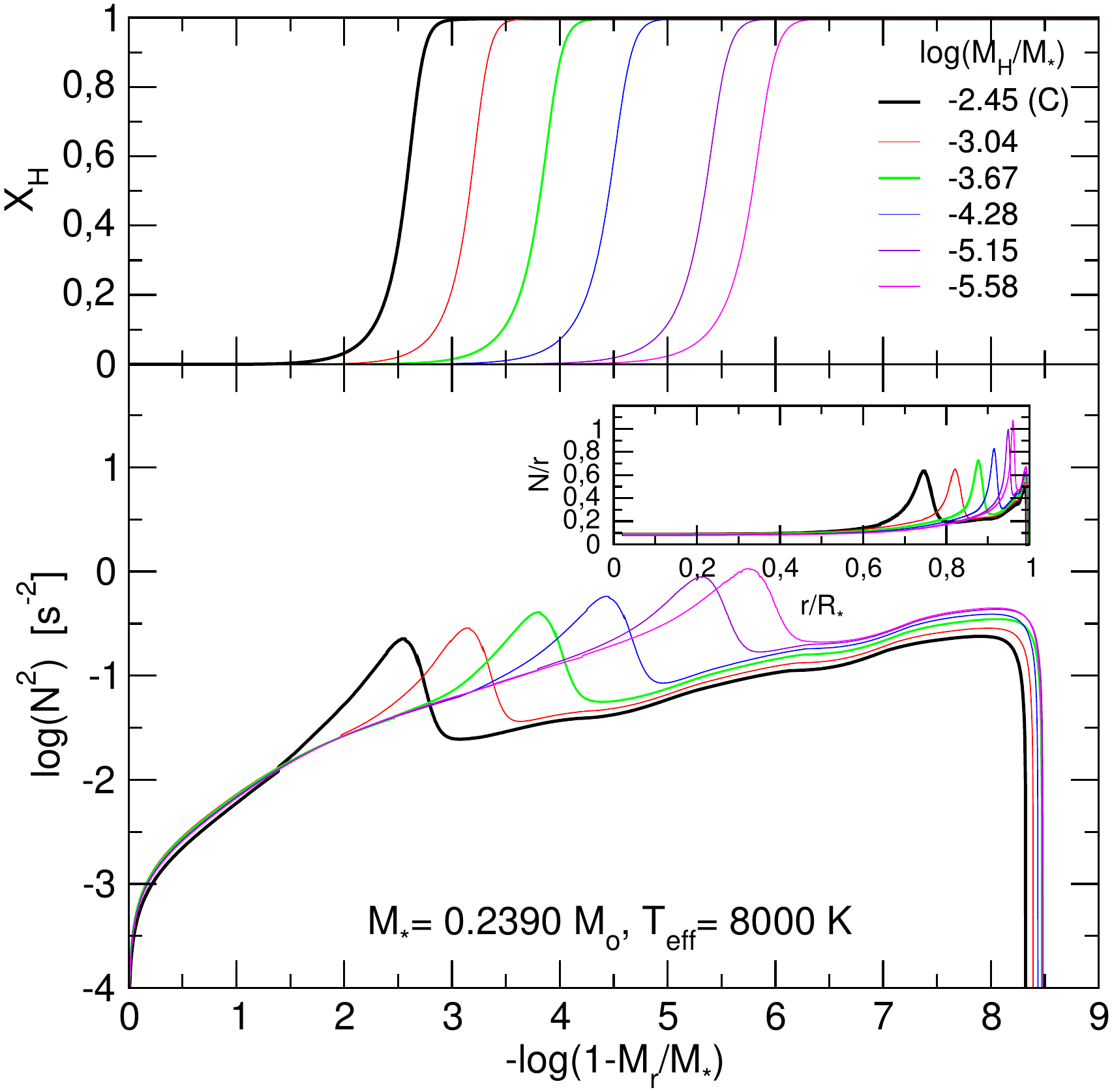} 
\caption{Same as in Fig. \ref{figure_02}, but for WD models with
  $M_{\star}= 0.2390 M_{\sun}$.}
\label{figure_03} 
\end{center}
\end{figure}

In the upper panels of Figs. \ref{figure_02} and \ref{figure_03}
we show the internal chemical profiles for H
corresponding to WD models at $T_{\rm eff} \sim 8000$ K with
$M_{\star}= 0.1554 M_{\sun}$ and 
$M_{\star}= 0.2390 M_{\sun}$. In each case, we show the profile
corresponding to the canonical envelope with thick black line,
and the thin H envelopes with lines of
different colors. For the models
with $M_{\star}= 0.1554 M_{\sun}$ (Fig. \ref{figure_02}), we note that
the envelopes with
$\log(M_{\rm H}/M_{\star})= -2.50$ and $-3.00$ have a double-layered shape,
which consists  of a  pure H  envelope surrounding a layer rich in H
and He. In the other envelopes (included the canonical
envelope),
the transition regions are characterized by single-layered chemical
profiles. Details are explained in the next Section.
In the case of the models with
$M_{\star}= 0.2390 M_{\sun}$ (Fig. \ref{figure_03}), the He/H transition
region has a single-layered shape for all the H envelope thicknesses
considered. 

\section{Single- and double-layered chemical structure of the H envelope}
\label{single-double}

Here, we show that in the case of ELM WD models, that is, WD models
with stellar masses in the interval of $0.1554 \leq M_{\star}/M_{\sun}\leq
0.1762$, the H envelopes can have a chemical structure of double layer,
for certain interval of effective temperatures well within the instability
strip of ELMV stars. We adopt the sequence with
$M_{\star}= 0.1554 M_{\sun}$ as our test case and compute several
model sequences with H envelope thicknesses in the range $-5.20 \leq
\log(M_{\rm H}/M_{\star}) \leq -1.68$ with a small step $\Delta(M_{\rm
  H}/M_{\star})$.  This range of envelope thicknesses includes the
canonical value $M_{\rm H}= 2.09 \times 10^{-2}M_{\star}$ for this template
sequence of models ($M_{\star}= 0.1554 M_{\sun}$).

\begin{figure} 
\begin{center}
\includegraphics[clip,width=9 cm]{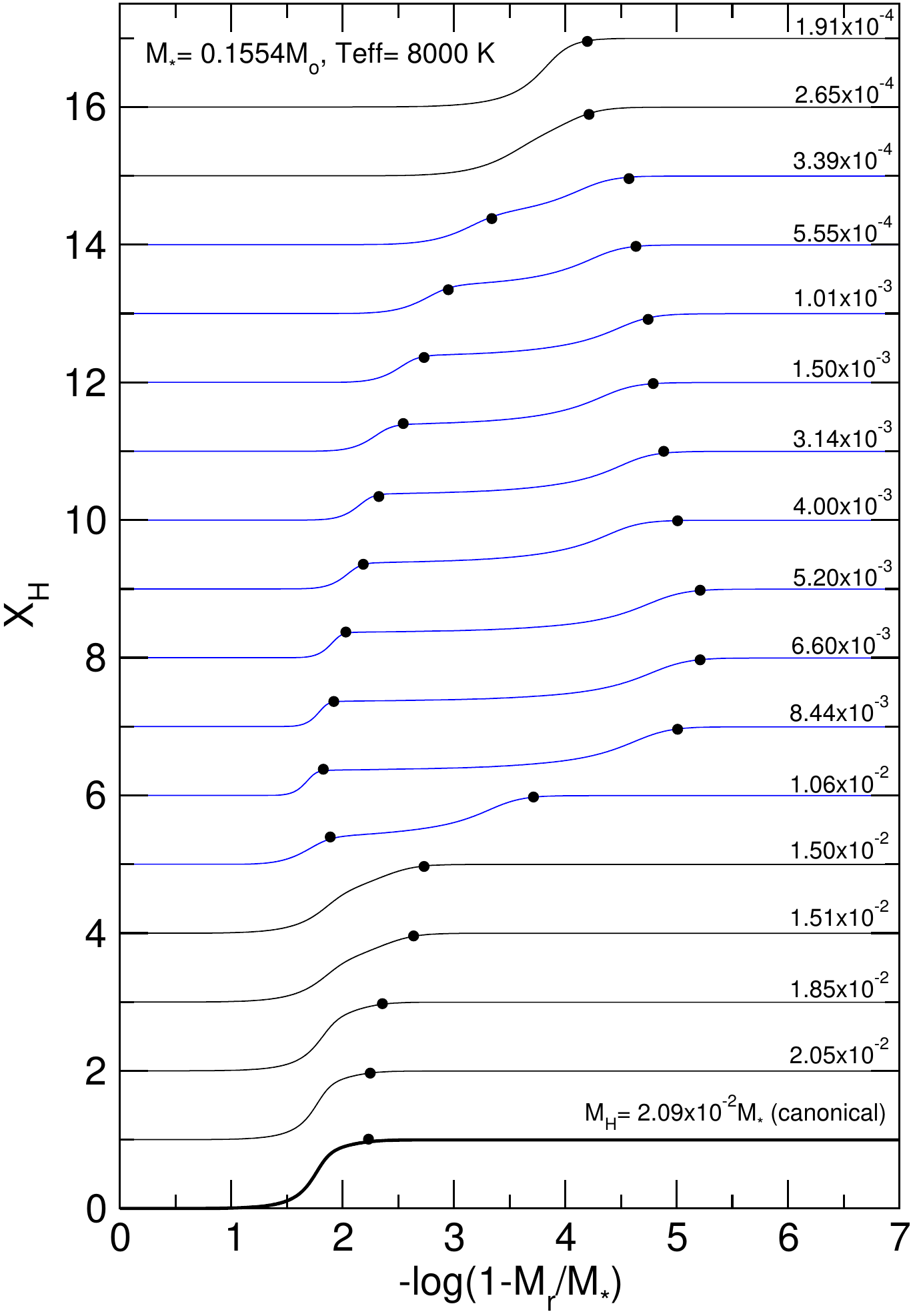} 
\caption{The fractional abundance of H ($X_{\rm H}$)  versus
  the outer mass fraction coordinate, for ELM WD models
  with $M_{\star}= 0.1554 M_{\sun}$ and $T_{\rm eff}= 8000$ K. The lowest
  curve (thick black) corresponds to a model characterized by
  the thickest (canonical) envelope for this stellar mass, while the upper curves
  (which have been artificially displaced upwards for clarity) correspond to
  models with H envelopes having decreasing thicknesses. The values of
  $M_{\rm H}$ are shown at the right of the figure. Black (blue) curves
  correspond to the case of H envelopes with single-layered (double-layered)
  chemical structures. Black dots mark the location of each step
  in the chemical profile.}
\label{figure_appendix} 
\end{center}
\end{figure}

In Fig. \ref{figure_appendix} we depict the fractional abundance of H
as a function of the outer mass fraction for models having $M_{\star}= 0.1554
M_{\sun}$ and $T_{\rm eff} \sim 8000$ K with a subset of H envelope
thicknesses. The curves have been displaced upwards arbitrarily for
clarity. We note that, for thick H envelopes, with values close to the
canonical one ($M_{\rm H} \gtrsim 1.50 \times 10^{-2}M_{\star}$), the
He/H chemical transition region has a single-layered structure at
$T_{\rm eff} \sim 8000$ K (lower black curves in
Fig. \ref{figure_appendix}). This is because such thick H
envelopes experience residual H nuclear burning, and this constitutes
the main energy source of the WD. This, in turn, results in very long
cooling timescales (of the
order of $\sim 10^9$ yr). Time-dependent element diffusion, acting during
these long cooling timescales, strongly changes  the  initial shape
of  the  H  and  He chemical  profiles  as  the  WD  cools,  forcing
He to  sink down and H  to  float  to  the surface.

When we consider slightly thinner H envelopes, nuclear burning is
much less important in relation to the cooling timescale of the WD model,
and the star cools
much faster. This being the case, the diffusion timescale at the basis
of the H envelope is longer than the cooling timescale of the WD. As a
result, during the cooling of the star, H floats to the surface at the
outer layers, but the basis of the H envelope remains virtually
unaltered. The consequence of this is that the envelope has a
chemical structure of double layer, consisting of a pure H envelope
surrounding a H- and He-rich shell. 

The presence of a double-layered chemical structure in the envelope of our models is a consequence of stellar evolution. A similar finding has been reported in
detail for models of DB WD stars \citep[e.g.][]{2004A&A...417.1115A}.
Fig. \ref{figure_appendix} shows that ELM WD models with $M_{\star}=
0.1554 M_{\sun}$ and $T_{\rm eff} \sim 8000$ K are expected to have a
double-layered chemical structure for H envelopes with thicknesses in
the range $3.4 \times 10^{-4}  \lesssim M_{\rm H}/M_{\star} \lesssim 1
\times 10^{-2}$ (blue curves in the Figure).
However, we realize that the existence of such double layers barely
impacts the pulsational properties of our ELM WDs.

Finally, for H envelopes even thinner ($M_{\rm H} \lesssim 3.4 \times
10^{-4}M_{\star}$), the H profile of models at the same $T_{\rm eff}$
adopts a single-layered chemical structure (upper black curves in
Fig. \ref{figure_appendix}). This is simply because for
ELMs with very thin H envelopes, the diffusion timescale at the tail of
the H distribution is extremely short (because of lower densities),
much shorter than the cooling times. Hence, the star evolves into a
structure with a single-layered chemical profile in a rather short period of
time.

In closing, it is worth mentioning that the effect described above for
the sequence of $M_{\star}= 0.1554 M_{\sun}$ is also verified for
more massive ELM WD model sequences ($0.1554 \leq M_{\star}/M_{\sun}\leq 0.1762$).

\section{The impact of thin H envelopes on the pulsation properties}
\label{pulsation-properties}

\begin{figure*} 
\begin{center}
\includegraphics[clip,width=15 cm]{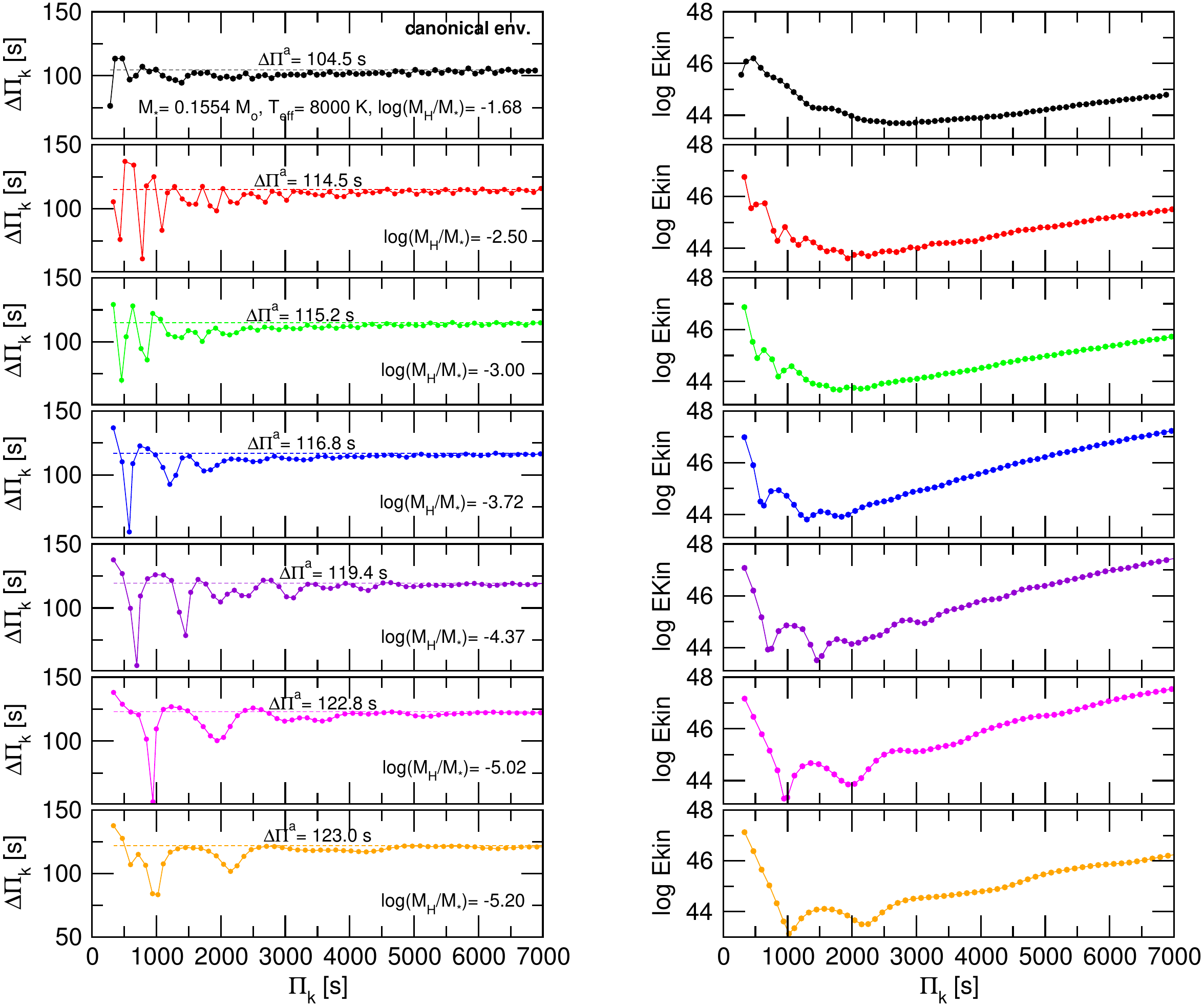} 
\caption{Left panels: the forward period spacing, $\Delta \Pi_k$ as a 
function of the pulsation periods, $\Pi_k$, for WD models with
  $M_{\star}= 0.1554 M_{\sun}$, $T_{\rm eff} \sim 8000$ K and
  different thicknesses of the H envelope (see  Table \ref{table_01} and
  Fig. \ref{figure_02}).  The upper panel corresponds to the WD
  model with canonical envelope.  The thin horizontal dashed lines
  correspond to the value of the asymptotic period spacing, $\Delta
  \Pi^{\rm a}$.  Right panels: the oscillation kinetic energy versus
  the periods for the same WD models shown in the left panel.}
\label{figure_04} 
\end{center}
\end{figure*}

The  shape of the chemical profiles leaves  notorious signatures in
the run of the squared critical frequencies, particularly, in the
Brunt-V\"ais\"al\"a frequency ($N$). In the lower panels of Figs.
\ref{figure_02} and \ref{figure_03} we show the logarithm of the
squared Brunt–V\"ais\"al\"a frequency for models with  $M_{\star}=
0.1554 M_{\sun}$ and  $M_{\star}= 0.2390 M_{\sun}$ ($T_{\rm
  eff} \sim 8000$ K).  There is a clear conection between the
chemical transition regions (upper panels) and the resulting features in
the run of the Brunt–V\"ais\"al\"a frequency for each model.

Let us now briefly examine the impact of the consideration of
thin H envelopes on the mode-trapping  properties of low-mass He-core
WD models. Mode trapping of $g$ modes in WDs
is a well-known mechanical
resonance for the mode propagation, that acts due to the
presence of chemical composition gradients  
\citep[see][for details]{1992ApJS...80..369B,1992ApJS...81..747B,
1993ApJ...406..661B,2002A&A...387..531C}.
Observationally, a possible indication of mode
trapping in a WD star is the departure from uniform period spacing.
According to the asymptotic theory of stellar pulsations, \emph{in absence of
chemical gradients} the pulsation periods of $g$ modes  with  high
radial  order  $k$  (long  periods) are expected to be uniformly
spaced with a constant period separation given by \citep{1990ApJS...72..335T}:

\begin{equation} 
\Delta \Pi_{\ell}^{\rm a}= \Pi_0 / \sqrt{\ell(\ell+1)},  
\label{aps}
\end{equation}

\noindent where

\begin{equation}
\Pi_0= 2 \pi^2 \left[\int_{r_1}^{r_2} \frac{N}{r} dr\right]^{-1}.
\label{p0}
\end{equation}

Actually, the period separation in \emph{chemically stratified} WD model
stars is not constant in anyway, except for very-high radial order modes. 
We define the forward period spacing as $\Delta \Pi_k= \Pi_{k+1}-\Pi_k$.
Stellar  models harboring a single chemical transition region (He/H) ---like those
considered here--- local minima in $\Delta \Pi_k$ are generally associated with
modes trapped in the H envelope, while  local maxima  in $\Delta
\Pi_k$ correspond to modes trapped in the core region.

\begin{figure*} 
\begin{center}
\includegraphics[clip,width=15 cm]{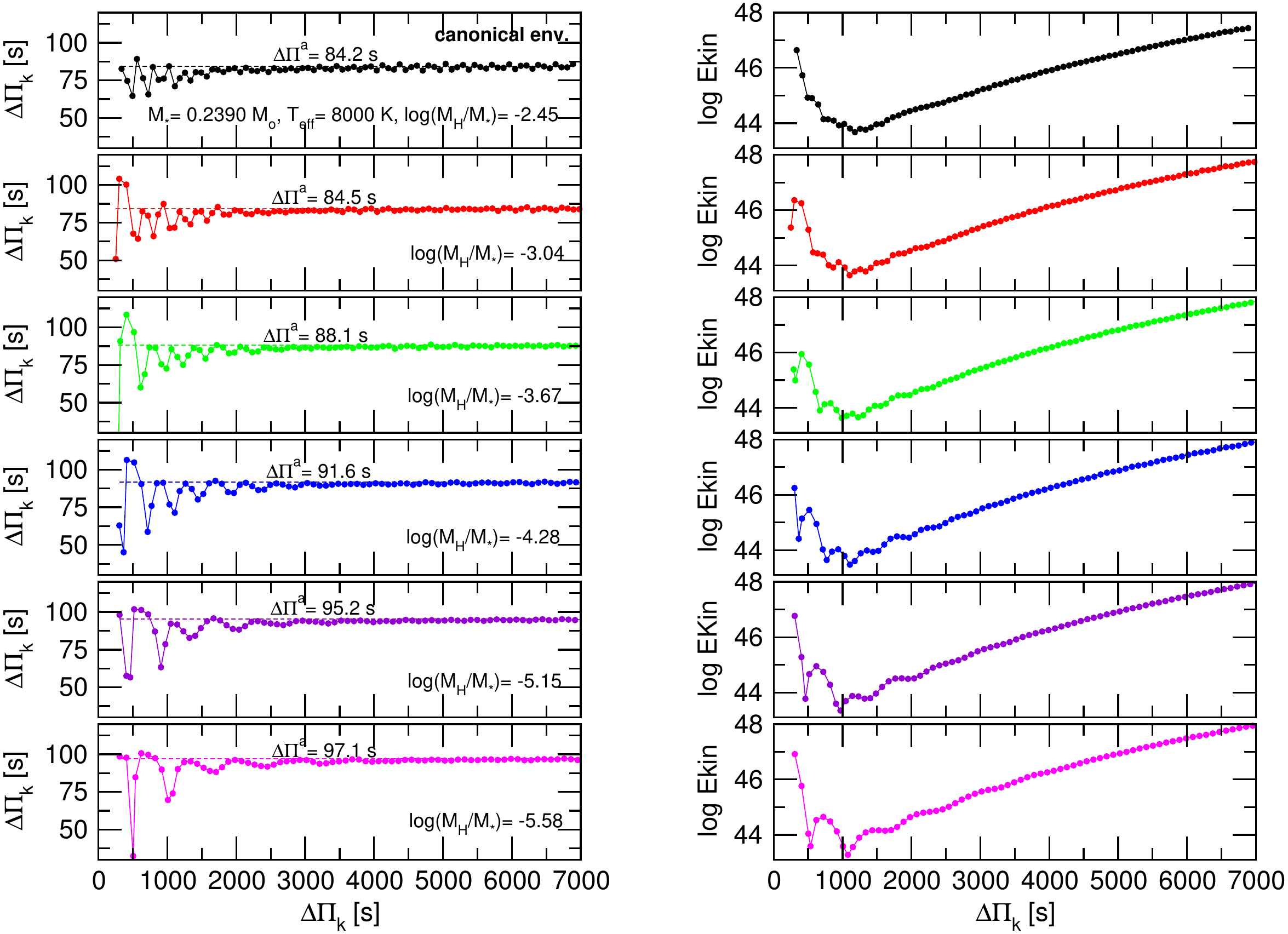} 
\caption{Same as in Fig. \ref{figure_04}, but for  WD models with
  $M_{\star}= 0.2390 M_{\sun}$ (see Table \ref{table_01} and
  Fig. \ref{figure_03}).}
\label{figure_05} 
\end{center}
\end{figure*}

The left panels of Figs. \ref{figure_04} and \ref{figure_05}
show $\Pi_k - \Delta \Pi_k$ diagrams for the same WD models
depicted in Figs. \ref{figure_03} and \ref{figure_04}. These models
are characterized by $M_{\star}= 0.1554 M_{\sun}$ and
$M_{\star}= 0.2390 M_{\sun}$ at $T_{\rm eff} \sim 8000$ K, and
different thicknesses of the H envelope. In each panel, the horizontal
dashed lines correspond to the asymptotic period spacing. Models with decreasing H envelope thicknesses
are displayed from top to bottom,
starting with the case of the canonical envelope.  By examining
the plots, several aspects are worth mentioning. To begin with, the asymptotic
period spacing increases for decreasing H envelope thickness. This is
because the integral in Eq. (\ref{p0}) for the quantity $\Pi_0$
is smaller for thinner H
envelopes, by virtue that the bump in the Brunt-V\"ais\"al\"a
frequency due to the He/H chemical interface becomes progressively narrow
in the radial coordinate $r$ as this interface is located
at more external layers. This can be clearly appreciated in the insets
of the lower panels of Figs. \ref{figure_02} and \ref{figure_03}, in which
we plot the quantity $N/r$ in terms of $r$. Since $\Pi_0$ is larger for
thinner H envelopes, the asymptotic period spacing increases (Eq. \ref{aps}).
In the case of the $0.1554 M_{\sun}$ models, we found that
$\Delta \Pi_{\ell}^{\rm a}$ experiences an increase of $15-18 \%$ when
we go from the canonical envelope ($\log(M_{\rm H}/M_{\star})= -1.68$)
to the thinest envelope ($\log(M_{\rm H}/M_{\star})= -5.20$) for this sequence.
For models with $M_{\star}= 0.2390 M_{\sun}$,
the variation (increase) of $\Delta \Pi_{\ell}^{\rm a}$ amounts
to $13-15 \%$ from the canonical envelope ($\log(M_{\rm H}/M_{\star})= -2.45$)
to the thinest one ($\log(M_{\rm H}/M_{\star})= -5.58$) for this sequence.

Another outstanding feature to be noted from the left panels of
Figs. \ref{figure_04} and \ref{figure_05} is connected with the changes in the
mode-trapping properties when we consider H envelopes progressively thinner.
Indeed, we note that for thick envelopes, including the canonical one,
the period-spacing distribution of $g$ modes exhibits 
a regular pattern of mode trapping with a very short trapping 
cycle ---the $k$ interval ($\Delta k$) between two trapped
modes. For instance, in the case of the $0.1554 M_{\sun}$
models, we found a trapping cycle of $\Delta k \sim 1-3$ for H envelope
thicknesses in the range $-3 \lesssim \log(M_{\rm H}/M_{\star}) \lesssim -1.7$.
When we consider thinner H envelopes, the trapping cycle and the trapping
amplitude increase. For instance, for $\log(M_{\rm H}/M_{\star})= -4.37$
we obtain $\Delta k \sim 5$, and for $\log(M_{\rm H}/M_{\star})= -5.20$
we have $\Delta k \sim 9$. A similar situation is found for the
models with $M_{\star}= 0.2390 M_{\sun}$ (Fig. \ref{figure_05}).

A common feature for all the values of $\log(M_{\rm H}/M_{\star})$
considered in both the $0.1554 M_{\sun}$ and
$0.2390 M_{\sun}$ sequences is that the mode-trapping signatures 
reflected by $\Delta \Pi_k$ vanish for very large radial orders
(very long periods), in which case $\Delta \Pi_k$
approaches to $\Delta \Pi_{\ell}^{\rm a}$, as predicted by
the asymptotic theory (see  Figs. \ref{figure_04} and
\ref{figure_05}). 

Mode-trapping effects also translate into local minima and maxima in the
kinetic energy of oscillation, $E_{\rm kin}$,  which generally
correspond to  modes  partially confined to  the core regions
and  modes  partially trapped in the envelope. This can be
appreciated in the right panels of Figs. \ref{figure_04} and
\ref{figure_05}. The behaviour described above for $\Delta \Pi_k$
is also found in the case of $E_{\rm kin}$, that is, the mode-trapping
cycle and amplitude increase with decreasing H envelope thickness.
Unfortunately,  the kinetic oscillation energy is a quantity
very difficult to estimate from observations alone.

\section{Asteroseismological analysis: period-to-period fits}
\label{astero}

\begin{table*}[t]
\centering
\caption{Stellar  parameters derived using 1D and 3D model atmospheres of the ELMVs shown in this work.}
\begin{tabular}{ccccccc}
\hline 
\hline
\vspace{1mm}
Star & $T^{1D}_{\rm eff}$ & $\log(g)^{1D}$ & $M^{(1D)}_{\star}$ & $T^{3D}_{\rm eff}$ &  $\log(g)^{3D}$ & $M^{(3D)}_{\star}$  \\
       &  [K]            & [cgs]    &    [$M_{\sun}$] &   [K]          & [cgs]    &    [$M_{\sun}$]  \\
\hline
J1112  & $9590 \pm 140$ &  $6.36 \pm 0.06$  & 0.179$^{\rm a}$  & $9240 \pm 140$ & $6.17 \pm 0.06$  & 0.169$^{\rm b}$ \\ 
J1518  & $9900 \pm 140$ &  $6.80 \pm 0.05$  & 0.220$^{\rm a}$  & $9650 \pm 140$ & $6.68 \pm 0.05$  & 0.197$^{\rm b}$ \\ 
J1738  & $9130 \pm 140$ &  $6.55 \pm 0.06$  & 0.181$^{\rm c}$  & $8910 \pm 150$ & $6.30 \pm 0.10$  & 0.172$^{\rm b}$ \\
J1735  & ---            & ---               &   ---          & $7940 \pm 130$  & $5.76 \pm 0.08$ & 0.142$^{\rm d}$ \\
\hline
\end{tabular}

\label{tablacero}
{\footnotesize  Notes: 
$^{\rm a}$\citet{2013ApJ...765..102H}.
$^{\rm b}$ Determined using the corrections for 3D effects by \citet{2015ApJ...809..148T}.}
$^{\rm c}$\citet{2015MNRAS.446L..26K}.  
$^{\rm d}$\citet{2017ApJ...835..180B}.
\end{table*}

We perform asteroseismological period-to-period fits 
to the complete set of ELMV stars known up to date\footnote{Except for
J1343+0826 \citep{2018MNRAS.478..867P}
because according to  \cite{2018arXiv180504070P}, there is only one
detected period which is not enough to perform a period fit.} (both confirmed
and suspected). However, at variance with \cite{2017A&A...607A..33C},
in this paper we only show the results for four out of nine ELMVs, following
the suggestion of our referee. The reason is that the remainder five stars
show very few periods
($\leq 3$) and/or large uncertainty in one or more of them.
The spectroscopic parameters for the ELMVs shown in this work,
for the 1D and 3D model atmosphere along with their uncertainties,
is displayed in Table~\ref{tablacero}. As in \cite{2017A&A...607A..33C}, we search for
a model that best matches the individual pulsation periods of the star
under analysis. The  quality of  the  match  between the   theoretical
pulsation  periods ($\Pi_k^{\rm  T}$) and  the observed   individual  periods
($\Pi_i^{\rm  O}$) is assessed by computing a merit function, which is defined
as: 

\begin{equation}
\label{chi}
\chi^2(M_{\star},  T_{\rm   eff}, M_{\rm H})=   \frac{1}{n} \sum_{i=1}^{n}   \min[(\Pi_i^{\rm   O}-   \Pi_k^{\rm  T})^2], 
\end{equation}

\noindent where $n$ is the number of observed periods. The ELM model having
the lowest value of $\chi^2$, if exists, is adopted as the ``best-fit
model''.  We compute the merit function
$\chi^2=\chi^2(M_{\star}, T_{\rm eff}, M_{\rm H})$ for our set of stellar masses ($0.1554$,
$0.1612$, $0.1650$, $0.1706$, $0.1762$, $0.1805$, $0.1863$, $0.1921$,
$0.2025$,  $0.2390$,  $0.2707$,  $0.3205$,  $0.3624$, and $0.4352 \ M_{\sun}$),
covering a wide range in effective temperature
$13000 \gtrsim T_{\rm eff} \gtrsim 6000\ $ K and also considering the thickness of
the H envelope in the interval
 $-5.8 \lesssim \log(M_{\rm H}/M_{\star}) \lesssim -1.7$
 (depending on the stellar mass). This complete set
comprises a total of  $\sim 17000$ WD configurations. 

First, we consider that  all of the observed periods
are associated with $\ell= 1$ $g$ modes, and we take the
set of observed periods,  $\Pi_i^{\rm  O}$, of each star into account
in order to assess
the quality function given by Eq.~(\ref{chi}). Next, we consider
a mix of $g$ modes associated with both $\ell= 1$ and $\ell=2$.
Because we generally do not find suitable solutions for $\ell= 1$ only, 
we display the cases for $\ell= 1$ and $\ell= 2$ combined with only two
exceptions. Figures~\ref{fig:j1112} to \ref{fig:j1735} show the
projection on the effective temperature versus the stellar mass plane
of the inverse of the quality function, $(\chi^2)^{-1}$, for each
ELMV under consideration, taking the corresponding set
of observed periods into account,
analogously to \cite{2017A&A...607A..33C}.
We include the effective temperatures and the stellar
masses of every ELMV  along with their uncertainties for the 1D (orange
box) and 3D \citep[][green box]{2015ApJ...809..148T}
model atmosphere  determinations. The
uncertainty considered for all the stellar masses is a $15\%$ of the
total mass. This is the characteristic difference in
the value of the mass as derived from independent sets of evolutionary tracks
\citep[see][]{2017A&A...607A..33C}.
Each point ($M_{\star}$,$T_{\rm eff}$) in the maps corresponds to
an H envelope mass value ($M_{\rm H}/M_{\star}$) that maximizes the
value of $(\chi^2)^{-1}$ for that stellar mass and effective
temperature.  All ranges (for the cases with $\ell= 1, 2$)
have been adjusted so we can show the best
period fits in a region close to that of interest.
As already established, the value of $\chi^2$ indicates 
the goodness of the match between the observed and the theoretical
periods: the better the period match, the lower the value of $\chi^2$
---in the figures, the greater the value of $(\chi^2)^{-1}$, which is
shown by a color coding. If there is a single maximum for a given star,
we adopt the corresponding model as the asteroseismological solution.
Unfortunately, in the cases we study there are multiple possible
solutions, and then we are forced to apply an external constraint to
adopt one ---usually, the uncertainty in the
effective temperature, given by the spectroscopy and, at variance
with \cite{2017A&A...607A..33C}, we also employ
the constraint of the stellar mass as given by the spectroscopic
determinations.

\subsection{The case of SDSS J111215.82+111745.0}

\begin{figure} 
\begin{center}
  \includegraphics[clip,width=8 cm]{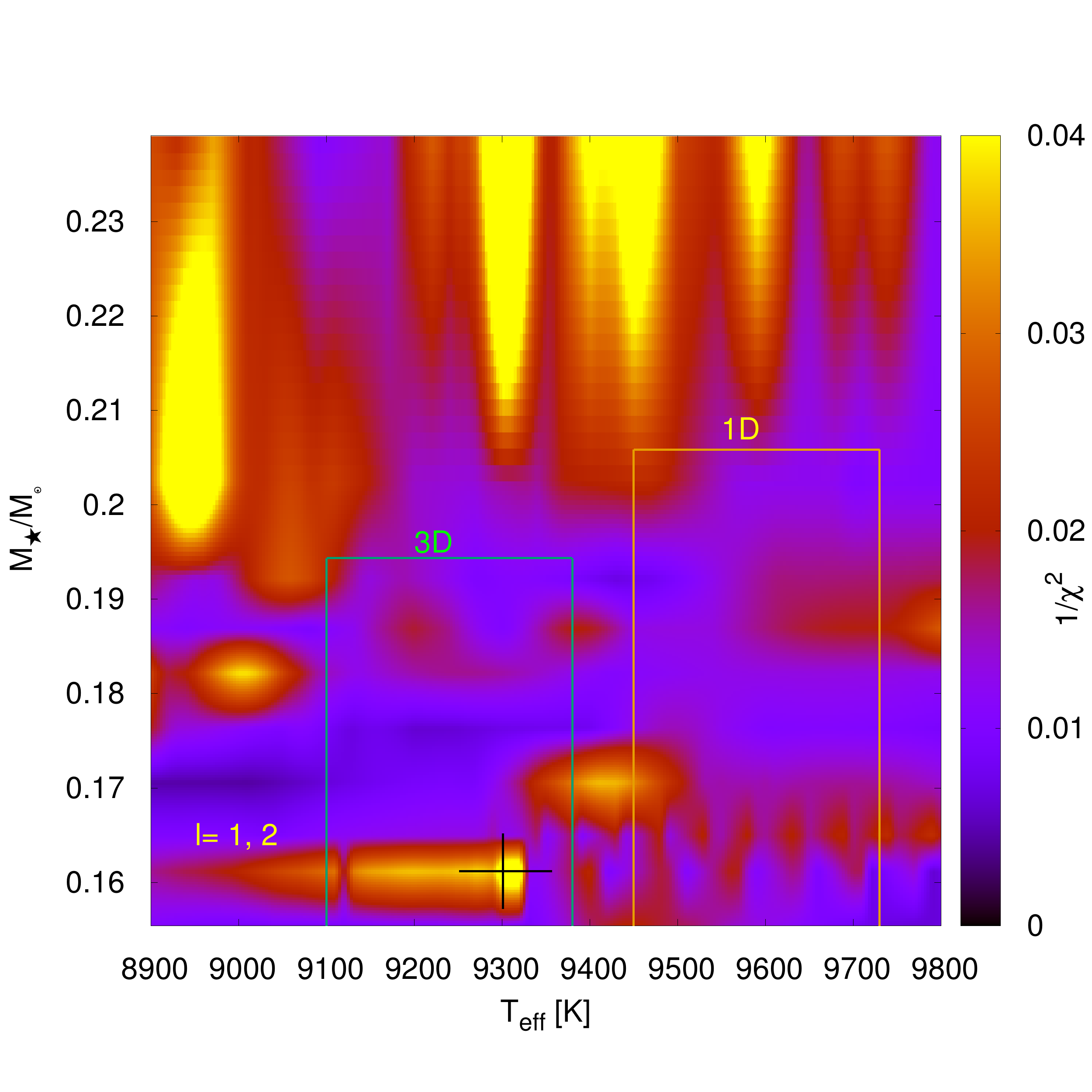}
    \caption{Projection on the effective temperature versus the stellar mass plane of the inverse of the quality function considering $\ell= 1, 2$ for the set of periods of J1112. The value of the thickness of the H envelope for each stellar mass corresponds to the sequence with the largest value of the inverse of the quality function for that stellar mass. The boxes depict the spectroscopic $T_{\rm  eff}$ and $M_{\star}$ determined for J1112 along with their uncertainties, for the 1D and 3D model atmosphere. The ranges taken in the three axes are focused on those of interest. The black cross marks the selected model.}
\label{fig:j1112} 
\end{center}
\end{figure}

SDSS J111215.82+111745.0 (hereafter J1112) exhibits a set of seven
independent periods,
according to \cite{2013ApJ...765..102H}. At variance with
\cite{2017A&A...607A..33C}, this time we only consider the set of the
five largest periods (see Table~\ref{tab:periods_comparison})
because the reality of the two shortest periods
($107.56$ and $134.275\ $s)
has to be confirmed. The $\ell= 1$ case has multiple possible
solutions and we can mention that there is one fit that lies within
the 1D box at $\sim 9670\ $K, for $0.1869 \ M_{\sun}$ and $\log(M_{\rm
  H}/M_{\star})= -5.19$, with $(\chi^2)^{-1}= 0.014$. The case of
$\ell= 1, 2$ shows the best period fit at a $T_{\rm eff}$ much lower
than expected ($\sim 8660\ $K, for $0.1650 \ M_{\sun}$ and
$\log(M_{\rm H}/M_{\star})= -1.82$ (canonical), with $(\chi^2)^{-1}=
0.38$). However, if we focus on ranges closer to the values allowed by
the spectroscopy, as we show in Fig.~\ref{fig:j1112}, we find two
possible solutions, one at $\sim 9301\ $K, for $0.1612 \ M_{\sun}$ and
$\log(M_{\rm H}/M_{\star})= -1.76$ (canonical) with $(\chi^2)^{-1}=
0.05$, and another one at $\sim 9406\ $K, for $0.1706 \ M_{\sun}$ and
$\log(M_{\rm H}/M_{\star})= -2.48$, with $(\chi^2)^{-1}= 0.037$. We
may choose the former although we note that these are not
associated with very good
period fits because they have low values of $(\chi^2)^{-1}$ in
comparison with the best solution.

For the purpose of establishing the accordance between the
observed and theoretical periods, we assess the absolute period differences
defined as $|\delta\Pi|= |\Pi^{\rm O}-\Pi^{\rm T}|$. We display in
Table~\ref{tab:periods_comparison} the corresponding
results for J1112 for $\ell=1, 2$, where  we indicate (column 6)
the value of the linear nonadiabatic
growth rate, $\eta$ ($\eta \equiv-\Im(\sigma)/ \Re(\sigma)$,
being $\Re(\sigma)$ and
$\Im(\sigma)$ the real and
the imaginary part, respectively, of the complex eigenfrequency
$\sigma$ computed  with the nonadiabatic version of the {\tt LP-PUL}
pulsation code  \citep{2006A&A...458..259C,2016A&A...585A...1C}). 
If $\eta$ is positive (negative), the mode is unstable (stable).
Unfortunately, in this case the nonadiabatic
analysis predicts that these periods are stable. 

\subsection{The case of SDSS J151826.68+065813.2}

\begin{figure} 
\begin{center}
\includegraphics[clip,width=8 cm]{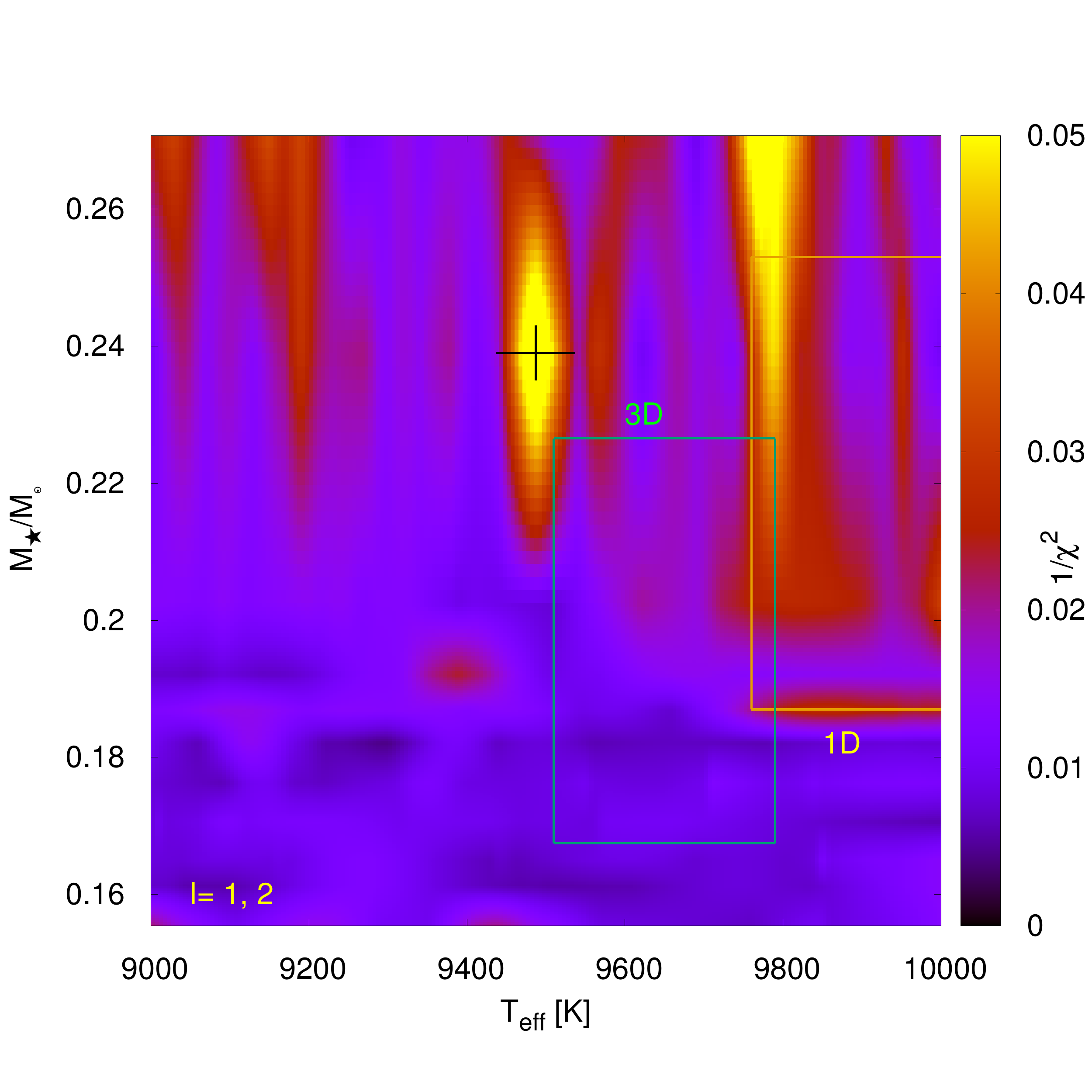} 
\caption{Same as Fig.~\ref{fig:j1112} but for J1518.}
\label{fig:j1518} 
\end{center}
\end{figure}

SDSS J151826.68+065813.2 (hereafter J1518) exhibits
seven periods (see Table~\ref{tab:periods_comparison}),
according to \cite{2013ApJ...765..102H}.
When we consider that these periods are associated with $\ell= 1$,
we find a possible solution at $\sim 9916\ $K, for $0.1762 \ M_{\sun}$ and
$\log(M_{\rm H}/M_{\star})= -2.5$, with $(\chi^2)^{-1}= 0.007$, close
to the 1D box. For the $\ell= 1, 2$ case, we find the best period fit at a
value of $M_{\star}$ higher than expected ($0.4352 \ M_{\sun}$, at $\sim 9717\ $K,
$\log(M_{\rm H}/M_{\star})= -3.69$, with $(\chi^2)^{-1}= 0.13$).
If we look closer to the ranges allowed by
spectroscopy as displayed in Fig.~\ref{fig:j1518}, we do not find any
solutions within the boxes but there is a good period fit lying close,
at $\sim 9487\ $K, characterized by $0.2390 \ M_{\sun}$,
$\log(M_{\rm H}/M_{\star})= -3.67$ and $(\chi^2)^{-1}= 0.07$. It represents the
best period fit in the ranges shown and we may adopt it as a
solution. As in the previous case, we display in Table~\ref{tab:periods_comparison}
the difference between the observed and the theoretical periods for the model
we adopt.

\subsection{The case of PSR J1738+0333}

PSR J1738+0333 is a millisecond pulsar with an ELMV as
companion which we call  (for short) J1738. \cite{2018arXiv180603650K}
made a reanalysis of this star, where they included additional observations
to the first one presented in \cite{2015MNRAS.446L..26K}. They also performed
an asteroseismic analysis, and they found a range of possible solutions.
We here show that study with more detail and we also perform an additional
search for a period fit.
Considering the different set of periods, corresponding to two years of observation,
we have a set of three periods from the 2014 data set ($\Pi_i^{\rm  O}= 1788.6$,
$2633.4$ and $3154.2\ $s) and another set of three periods from the 2017 data set
($\Pi_i^{\rm  O}= 1833.9$, $3321.7$ and $4980.6\ $s), both
according to \cite{2018arXiv180603650K}\footnote{Note that the present results may differ
  to some extent when compared with those from \cite{2017A&A...607A..33C}.
  This is because in that work we employed the 2014 period set quoted by
\cite{2015MNRAS.446L..26K}, which is somewhat different from that derived
in \cite{2018arXiv180603650K}.}. We have performed three different
period fits, two considering the 2014 and the 2017 data set separately,
(employing the sets from \cite{2018arXiv180603650K}), and another one
considering the six periods combined.
This last analysis is feasible, because for pulsation measurements of
different epochs of observation there is a meaningful difference in both
the periods and their amplitudes,
that can be associated with different modes becoming visible, as already
established for ZZ Ceti stars, for instance, in \cite{1998ApJ...495..424K}.

\begin{figure*}[t]
  \begin{center}
    \subfigure[Set of 2014, 3 periods]{\label{fig:j1738-14-l1}\includegraphics[clip,width=5.8 cm]{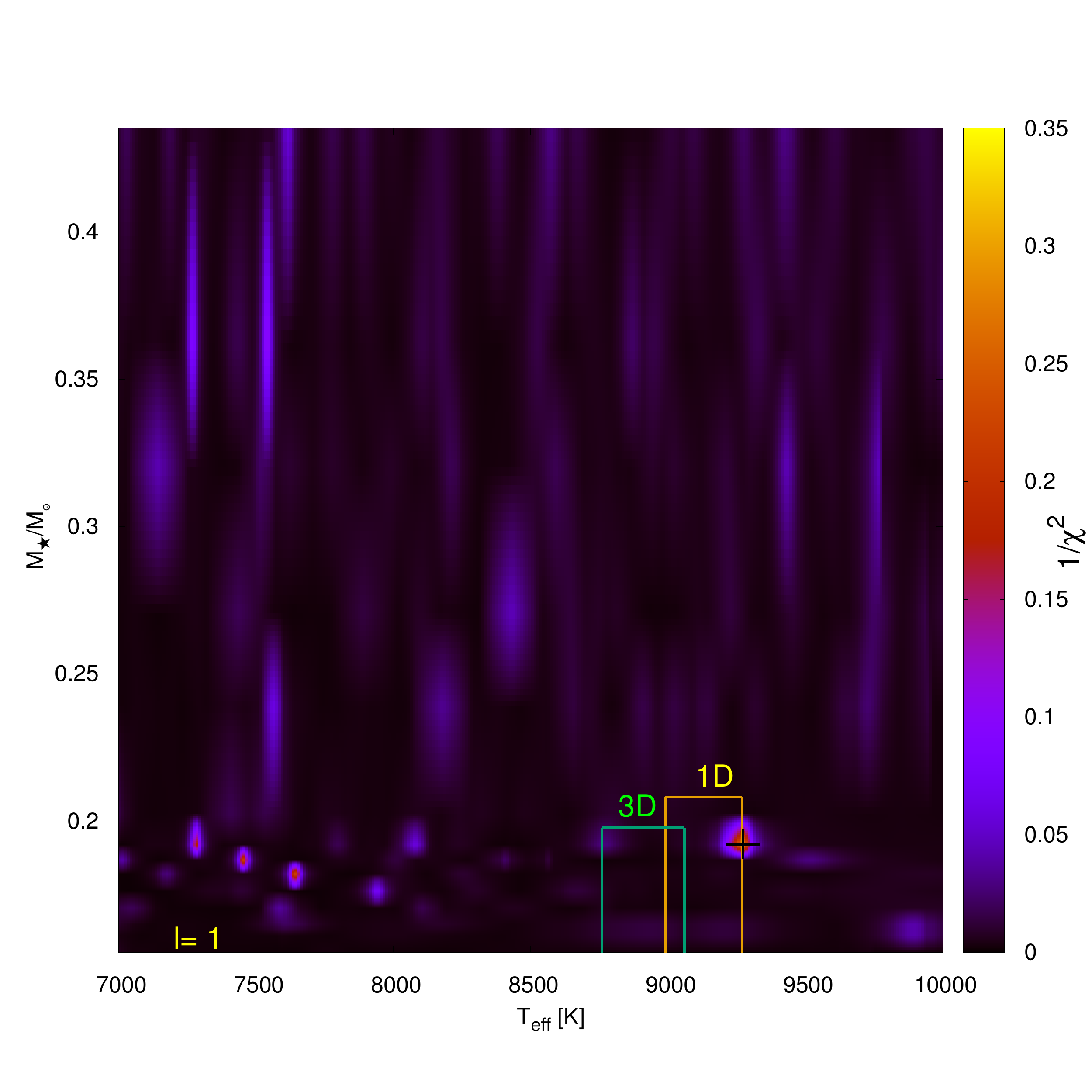}}
    \subfigure[Set of 2017, 3 periods]{\label{fig:j1738-17-l1}\includegraphics[clip,width=5.8 cm]{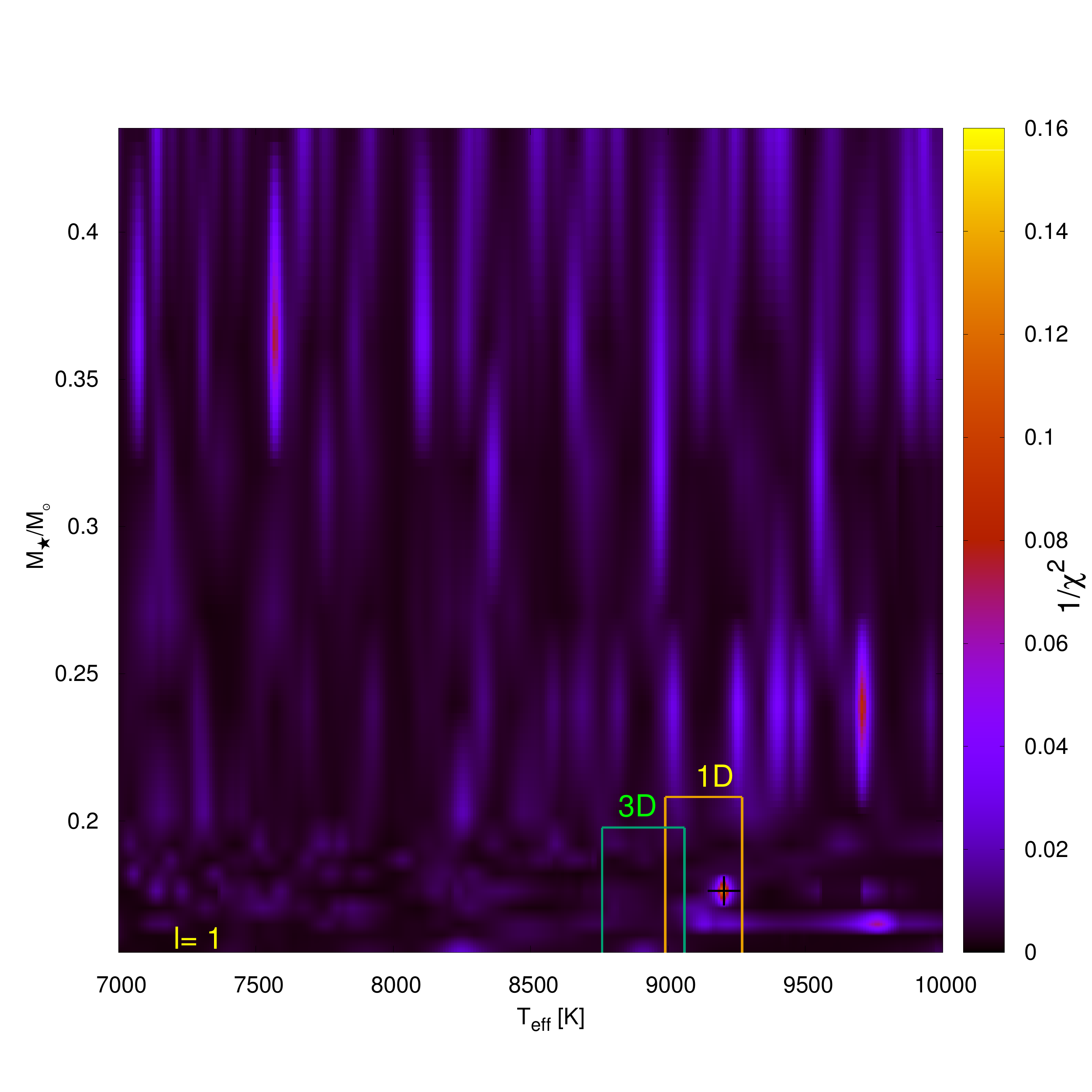}} 
  \end{center}
  \caption{Same as Fig.~\ref{fig:j1112} but for J1738,
  considering the periods of the 2014 data set, assuming they are associated
  with $\ell= 1$.}
\label{fig:j1738-l1} 
\end{figure*}

\begin{figure*}[t]
  \begin{center}
    \subfigure[Set of 2014, 3 periods]{\label{fig:j1738-14-l1l2}\includegraphics[clip,width=5.8 cm]{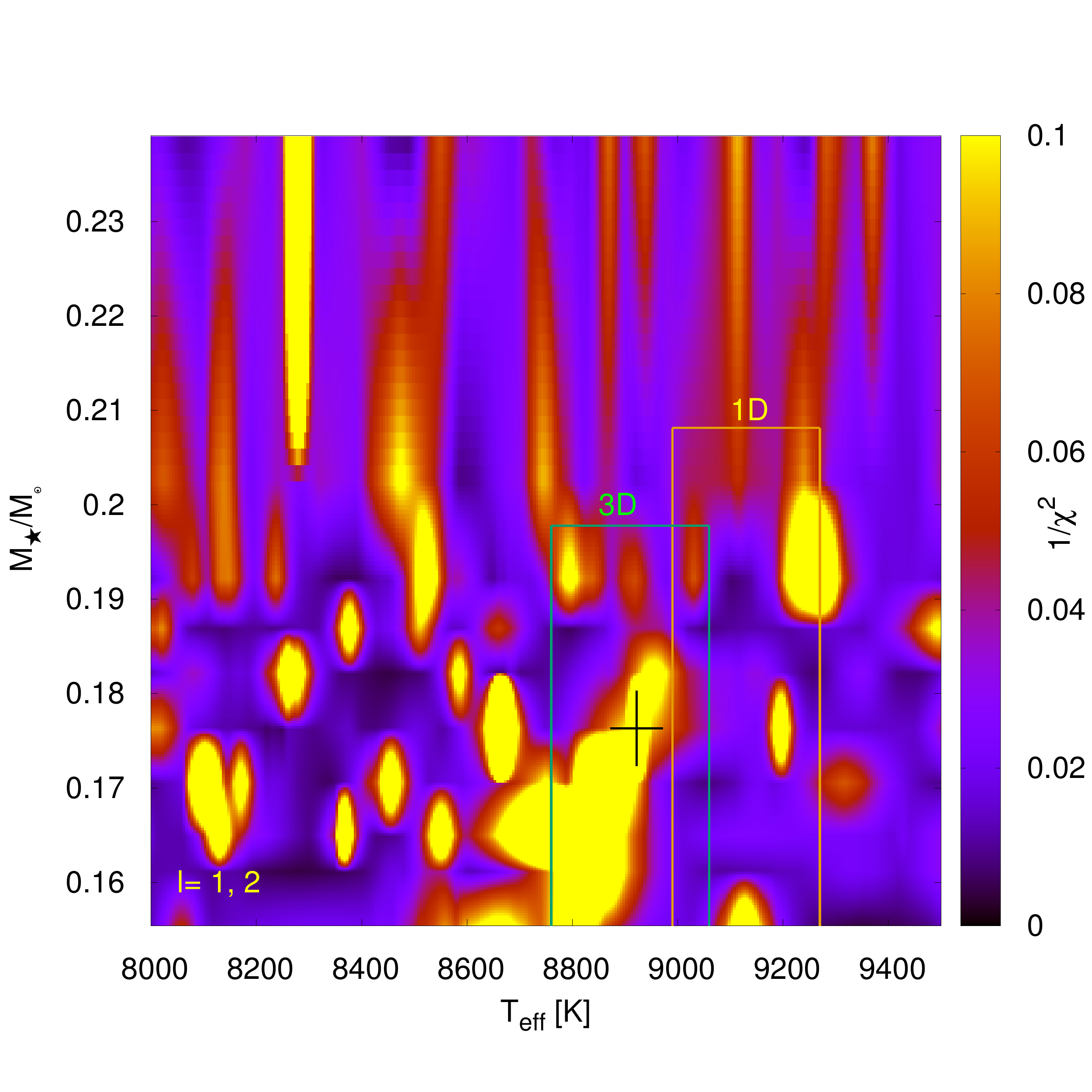}}
    \subfigure[Set of 2017, 3 periods]{\label{fig:j1738-17-l1l2}\includegraphics[clip,width=5.8 cm]{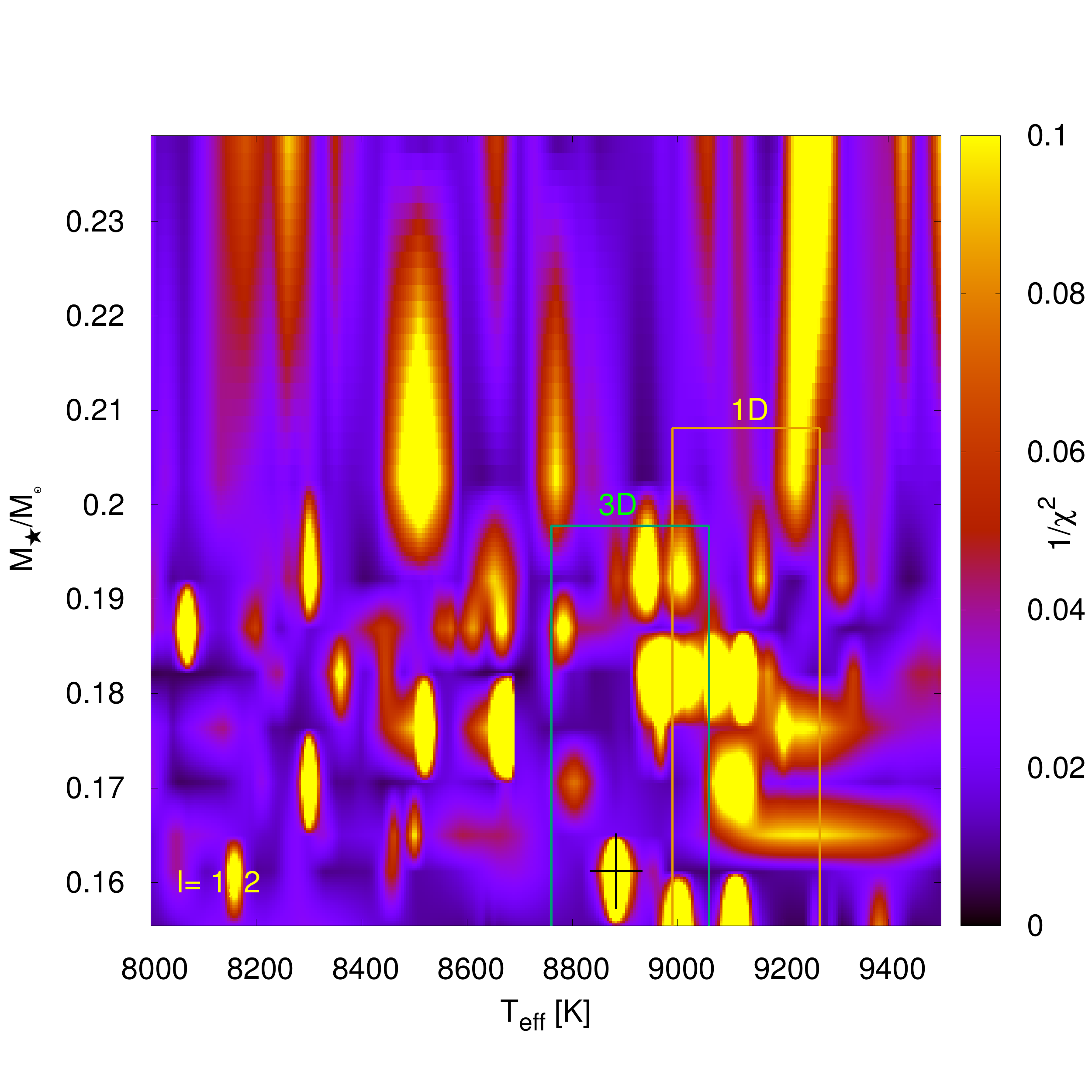}} 
    \subfigure[Set of 2014+2017, 6 periods]{\label{fig:j1738-1417-l1l2}\includegraphics[clip,width=5.8 cm]{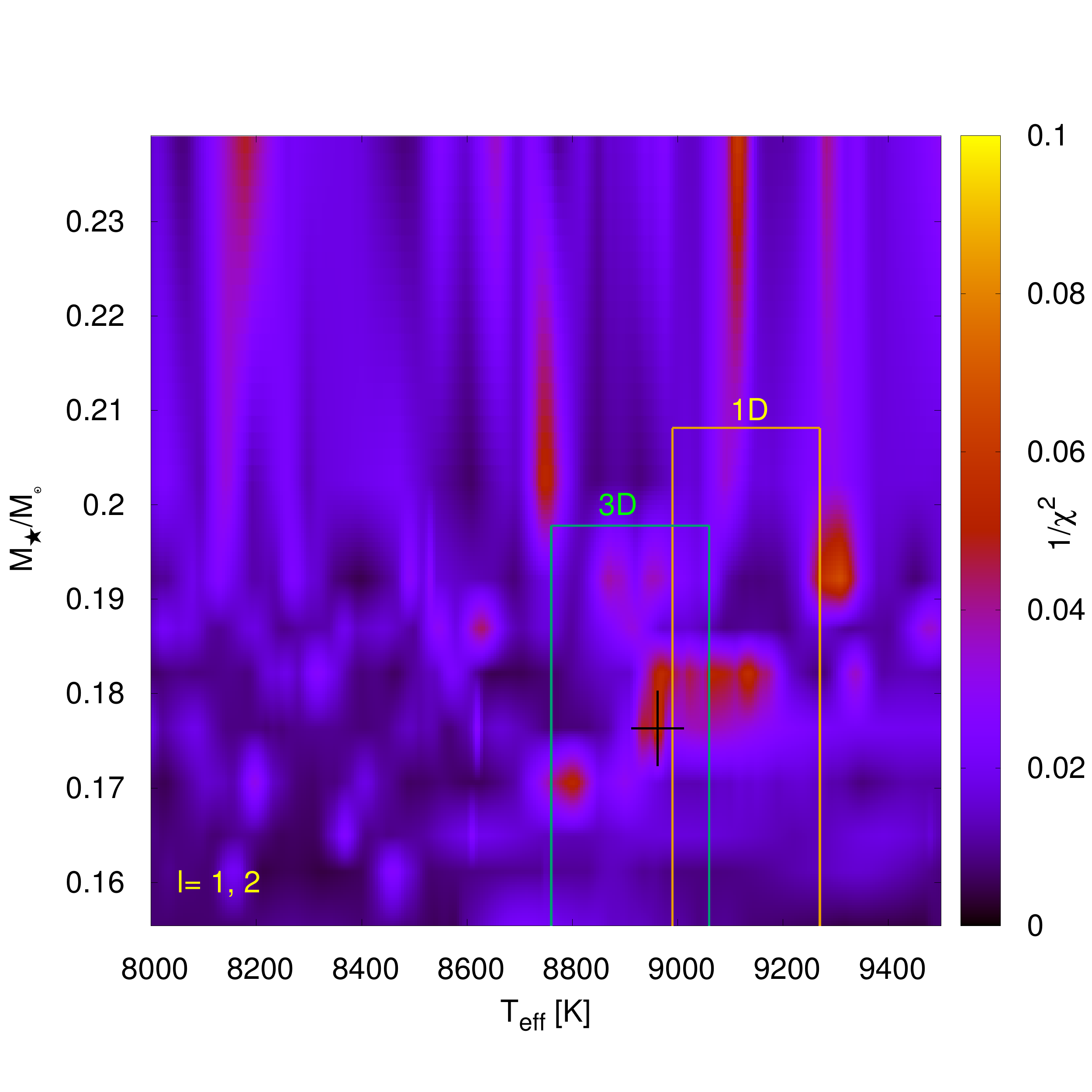}}
  \end{center}
  \caption{Same as Fig.~\ref{fig:j1112} but for J1738, for the set of periods corresponding to 2014, 2017 and 2014+2017.}
\label{fig:j1738}
\end{figure*}

When we consider the first case (2014) for $\ell= 1$, the best period fit 
lies at $9273\ $K, with $0.1921\ M_{\sun}$, $\log(M_{\rm H}/M_{\star})= -3.02$
and $(\chi^2)^{-1}= 0.31$, that is, almost within the 1D spectroscopic box and
it represents a very good solution. We show this in Fig.~\ref{fig:j1738-14-l1}.
For $\ell= 1, 2$, the best one lies at $9689\ $K, with $0.4352\ M_{\sun}$,
$\log(M_{\rm H}/M_{\star})= -3.21$ (canonical) and $(\chi^2)^{-1}= 13$, but it has
a stellar mass larger than expected. However, as shown in
Fig.~\ref{fig:j1738-14-l1l2}, there are many other solutions, for instance within
the 3D box, there is a possible solution at $8922\ $K, for $0.1762\ M_{\sun}$,
$\log(M_{\rm H}/M_{\star})= -5.42$ with $(\chi^2)^{-1}= 0.72$. 

When we consider the second case (2017) for $\ell= 1$, we find the best
period fit at $9204\ $K, for $0.1762\ M_{\sun}$ and
$\log(M_{\rm H}/M_{\star})= -5.43$, with $(\chi^2)^{-1}= 0.16$, that is, within
the 1D spectroscopic box. We show this case in Fig.~\ref{fig:j1738-17-l1}. For
$\ell= 1, 2$, the best one lies at $8554\ $K, for $0.1762\ M_{\sun}$,
$\log(M_{\rm H}/M_{\star})= -1.95$ (canonical), with $(\chi^2)^{-1}= 1.31$, but it
has a low value of $T_{\rm eff}$. We narrow down the ranges and
we show it in Fig.~\ref{fig:j1738-17-l1l2}. Once again, there are many possible
solutions. For instance, there is one at $8883\ $K, for
$0.1612\ M_{\sun}$ and $\log(M_{\rm H}/M_{\star})= -1.76$
(canonical), with $(\chi^2)^{-1}= 1.22$, which is a very good fit lying within
the 3D box.

When we consider the 2014+2017 data combined, we find for $\ell= 1$ a possible
solution at $8863\ $K, for $0.1921\ M_{\sun}$,
$\log(M_{\rm H}/M_{\star})= -2.35$ (canonical), with $(\chi^2)^{-1}= 0.006$ (not
shown). For $\ell= 1, 2$, we find the best solution at $9891\ $K, for
$0.4352\ M_{\sun}$ and $\log(M_{\rm H}/M_{\star})= -3.70$, with
$(\chi^2)^{-1}= 0.18$. Focusing on more appropriate ranges, as shown in
Fig.~\ref{fig:j1738-1417-l1l2}, we see that there is a good period fit lying
close to the 1D box, at
$9311\ $K, with $0.1921\ M_{\sun}$, $\log(M_{\rm H}/M_{\star})= -3.02$,
with $(\chi^2)^{-1}= 0.068$. Within the spectroscopic
boxes, we find a possible solution lying at $8962\ $K, for $0.1762 M_{\sun}$,
$\log(M_{\rm H}/M_{\star})= -1.95$ (canonical), with $(\chi^2)^{-1}= 0.062$.

From these results we can only conclude that the solutions have a stellar mass
in the interval of $M_{\star}= 0.1612 - 0.1921\ M_{\sun}$, with a constrained
$T_{\rm eff}$ in the range of $\sim 8883-9273\ $K, and an H envelope very poorly
constrained in the range of
$M_{\rm H}/M_{\star}= 3.75 \times 10^{-6} -1.74\times 10^{-2}$.
As expected, comparing Figs.~\ref{fig:j1738-14-l1l2} and
\ref{fig:j1738-17-l1l2} with \ref{fig:j1738-1417-l1l2}, we obtain less
possible solutions when increasing the
number of considered periods. Despite of this, the results obtained are not more
conclusive (the solutions do not have larger values of $(\chi^2)^{-1}$).

\subsection{The case of SDSS J1735+2134}

\begin{figure} 
\begin{center}
\includegraphics[clip,width=8 cm]{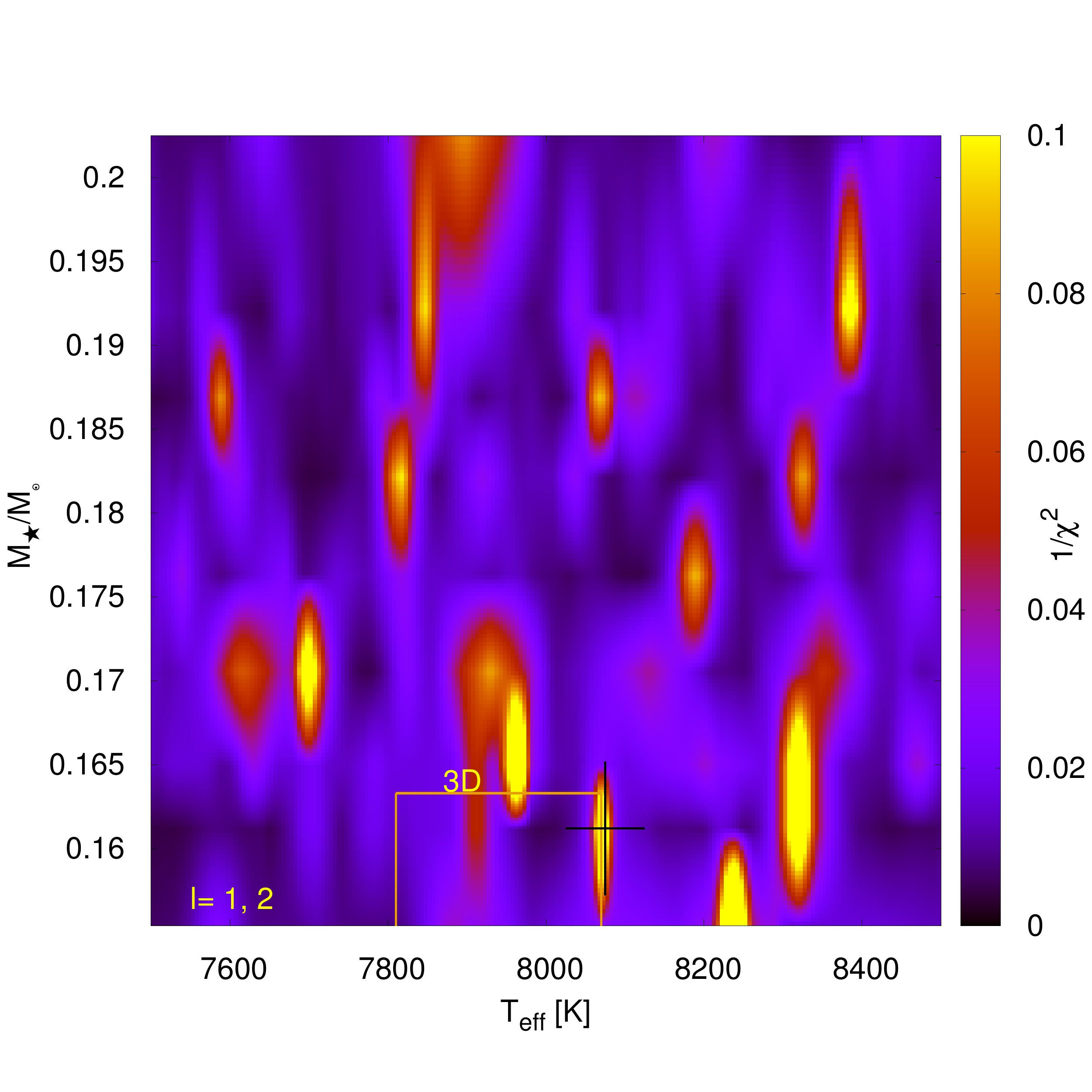} 
\caption{Same as Fig.~\ref{fig:j1112} but for J1735.}
\label{fig:j1735} 
\end{center}
\end{figure}

SDSS J1735+2134 (hereafter J1735) shows four independent periods
(see Table~\ref{tab:periods_comparison}), according
to \cite{2017ApJ...835..180B}.
The case of $\ell= 1$ shows a poor solution within the box at $8082\ $K, for
$0.1612 \ M_{\sun}$, and $\log(M_{\rm H}/M_{\star})= -1.76$ (canonical)
with $(\chi^2)^{-1}= 0.007$, while for the case of $\ell= 1, 2$, the period
fit with the best value is associated with a model consisting of
$0.2390 \ M_{\sun}$ (at $9892\ $K and
$\log(M_{\rm H}/M_{\star})= -4.28$, with $(\chi^2)^{-1}= 0.53$), once again,
outside the allowed ranges. If we
focus in the ranges allowed by spectroscopy as in Fig.~\ref{fig:j1735},
we can see good period fits close to the box at $7963\ $K, for $0.1650 \ M_{\sun}$
and $\log(M_{\rm H}/M_{\star})= -1.82$ (canonical), with $(\chi^2)^{-1}= 0.23$
and another at $8075\ $K, for $0.1612 \ M_{\sun}$
and $\log(M_{\rm H}/M_{\star})= -1.76$ (canonical), with $(\chi^2)^{-1}= 0.22$.
This last period fit may be choosen as a solution, and it is the same result
obtained in \cite{2017A&A...607A..33C}. We display the comparison between
theoretical and observed periods in Table~\ref{tab:periods_comparison}.

\begin{table*}[t]
\centering
\caption{Observed and theoretical periods  ($\ell= 1, 2$)  for the asteroseismological models for J1112 with ($M_{\star}$,$T_{\rm eff}$,$\log(M_{\rm H}/M_{\star})$) = ($0.1612\ M_{\sun}$, $9301\ $ K, $-1.76$), for J1518 with ($0.2390\ M_{\sun}$, $9487$ K, $-3.67$) and for J1735 with ($0.1612\ M_{\sun}$, $8075\ $K, $-1.76$). The harmonic degree $\ell$, the radial order $k$, the absolute period difference, and the nonadiabatic growth rate for each theoretical period are also displayed.}
\begin{tabular}{ccccccccc}
\hline
\hline
Star & $\Pi^{\rm O}$[s] & $\Pi^{\rm T}$[s] &  $\ell$ & $k$ & $|\delta\Pi|$[s] & $\eta[10^{-4}]$ &
Remark\\
\noalign{\smallskip}
\hline
J1112 &$1792.905 \pm 0.005$& $1797.822$& $1$ & $17$&  $4.92$ & $-0.00004$ & stable \\
      &$1884.599 \pm 0.004$& $1886.039$& $2$ & $32$&  $1.44$ & $-0.0155$ & stable \\
      &$2258.528 \pm 0.003$& $2266.323$& $2$ & $39$&  $7.79$ & $-0.131$ & stable \\
      &$2539.695 \pm 0.005$& $2536.215$& $2$ & $44$&  $3.48$ & $-0.399$ & stable \\
      &$2855.728 \pm 0.01$& $2861.968$& $2$ & $50$&  $6.24$ & $-1.14$ & stable \\
\hline
J1518 &$1335.318 \pm 0.003$& $1329.599$& $2$ & $28$&  $5.719$ & $0.463$ & unstable \\
      &$1956.361 \pm 0.003$& $1959.913$& $1$ & $24$&  $3.552$ & $0.653$ & unstable \\
      &$2134.027 \pm 0.004$& $2131.306$& $2$ & $46$&  $2.721$ & $0.504$ & unstable \\
      &$2268.203 \pm 0.004$& $2266.188$& $1$ & $28$&  $2.015$ & $0.766$ & unstable \\
      &$2714.306 \pm 0.003$& $2717.686$& $2$ & $59$&  $3.380$ & $-0.373$ & stable \\
      &$2799.087 \pm 0.005$& $2802.873$& $1$ & $35$&  $3.786$ & $1.14$ & unstable \\
      &$3848.201 \pm 0.009$& $3851.967$& $2$ & $84$&  $3.766$ & $-4.96$ & stable \\
\hline
J1735 &$3362.76 \pm 0.54$& $3359.87$& $2$ & $56$&  $2.89$ & $5.57$ & unstable \\
      &$3834.54 \pm 0.42$& $3831.65$& $2$ & $64$&  $2.89$ & $0.243$ & unstable \\
      &$4541.88 \pm 0.24$& $4542.92$& $2$ & $76$&  $1.04$ & $-14.3$ & stable \\
      &$4961.22 \pm 0.72$& $4960.70$& $1$ & $48$&  $0.52$ & $13.9$ & unstable \\
\hline
\end{tabular}
\label{tab:periods_comparison}
\end{table*}

\section{Summary and conclusions}
\label{conclusions}

In this work, we  have presented a thorough asteroseismological analysis
of pulsating ELM WD stars on the basis of our complete set
of fully evolutionary models that represent low-mass He-core WDs
harboring a range of H envelope thicknesses. This is the sixth paper in
a series of works dedicated to the study of pulsating low-mass He-core WDs
(including ELMVs). Here, we first explored the chemical profiles
of the grid of models having different H envelope thicknesses, and also
the impact on the adiabatic pulsation properties. Furthermore, we performed an 
asteroseismological analysis to four ELMV stars, those that show a
  sufficiently high number of periods and with small uncertainties to allow us
  to draw robust asteroseismological conclusions, in analogy to that of
  \cite{2017A&A...607A..33C}, but this time employing
this larger set of evolutionary sequences that expands the parameter space
by incorporating the thickness of the H envelope as a free parameter.

Next, we present a brief summary of the main results of this work:

\begin{itemize}
  
\item[-] The analysis of the internal chemical profiles of our ELM WDs
  models with thin H envelope puts in evidence the existence of a
  double-layered shape of the H envelope for some models lying well
  within the instability strip.

\item[-] When we considered thin H envelopes and analyzed the impact on
  the mode trapping properties of $g$ modes of our ELM WDs, we found that
  they strongly depend on $M_{\rm H}$. The period-spacing distribution
  acquires a steady form for thick envelopes (including the
  canonical one), with a very short trapping cycle, while for thinner H
  envelopes both the trapping cycle and the trapping amplitude increase.

\item[-] The asymptotic period spacing increases for decreasing H
  envelope thickness.
  
\item[-] The period-to-period fits show multiple solutions. Only with
  the inclusion of external
  constraints (i.e., spectroscopy values) we were able to adopt a model
  and in one case, we could only give a range of possible solutions. 
  The results are condensed in Table~\ref{tab:tablafinal}.
  
\item[-]  Some of the solutions found (see Table~\ref{tab:tablafinal})
  are characterized by thick (canonical) H envelopes and some by thin H envelopes.
  This result reinforces the findings of \cite{2018A&A...614A..49C} about the
  possible existence of ELM WDs with thin H envelope, thus leading to the 
  possibility that they could have been formed through
  unstable mass loss, for instance, via common-envelope episodes.
  
\item[-] Most of the solutions we adopted have a lower value of the
  stellar mass
  than those adopted in \cite{2017A&A...607A..33C}. The reason for this may be
  related to the fact that we are employing models with thinner H envelope
  and also because we are using a new constraint, the stellar mass.
  
\item[-] We generally did not find appropriate solutions considering modes associated
  only with $\ell= 1$, but with a mixture of $\ell= 1, 2$. Moreover, in most cases
  we found that the pulsation periods corresponding to the adopted asteroseismological
  models are associated with pulsationally unstable modes.
  
\end{itemize}

Given the scarcity of observed periods in the ELM WDs it is very
difficult to apply the methods of asteroseismology, specifically, to find
a unique solution compatible with the spectroscopic determinations. This
forces us to focus our exploration on the range of parameters (observational
boxes) dictated by the spectroscopy. In addition, when there is a change
in the periods determined for one star, the solutions change considerably.
Furthermore, only for one case, J1112, our non-adiabatic computations
predict that the adopted solutions are not pulsationally unstable. 

For the purpose of comparison with our previous work,
in Table~\ref{tab:comparison} we show the asteroseismological
results obtained in the present paper and in
\cite{2017A&A...607A..33C}. J1735 has not
been included in the Table because we obtain the exact same
solution when we allow the H-envelope thickness to vary.
It is interesting to note that,  when we incorporate
the thickness of the H envelope as a free parameter, for J1112
we still obtain a solution with a canonical H envelope, although it is a
different one because now we use the constraint of the
stellar mass to adopt a solution. On the whole, we can conclude
that the asteroseismological models in general change when the
H-envelope thickness is taken as a variable.
  
All in all, in this paper we have presented a thorough asteroseismological
study of ELMVs, using a complete set of fully
evolutionary models of He-core ELM WDs with different H envelope thicknesess.
In this way, we have pushed the limits of what is possible in terms
of deriving the internal structure of these stars through
asteroseismological period fits based on this grid. In order to
achieve progress in this field, and obtain
more robust asteroseismological solutions, it is necessary to obtain
richer observations of the pulsations of these stars. Indeed, due to
the few periods detected, we had to restrict ourselves to show the results
for stars with the richest period spectra and the less uncertainty in their
periods. The detection of a larger  number of pulsation periods of the known
ELMVs, and the discovery of new ELMV stars, will allow a substantial
progress in the knowledge of the internal structure of low mass WD
stars, the nature of  their predecessors, and the evolutionary
channels that lead to their origin \citep[see][]{2018A&A...614A..49C}.

\begin{table*}[t]
\centering
\caption{Main characteristics of the adopted 
  asteroseismological models and of the ranges of possible solutions
  for the ELMVs shown in this work.}
\begin{tabular}{lcccccc}
\hline
\hline
Star & $T_{\rm eff}$ [K]  & $\log(g)$ [cgs]  &  $M_{\star}[M_{\sun}]$ &  $\log(M_{\rm H}/M_{\star})$ & $\log(R_{\star}/R_{\sun})$ & $\log(L_{\star}/L_{\sun})$ \\
\hline
J1112$^{\rm *}$  & 9301  &     5.9695     & 0.1612          &-1.76          &-1.1623           &-1.4932 \\
J1518          & 9487  &     6.9994     & 0.2390          &-3.67           &-1.5916           &-2.3200 \\
J1738      &[8883,9273] &[6.0506,6.6923] &[0.1612,0.1921] &[-5.43,-1.76]   &[-1.5057,-1.2029] &[-2.2548,-1.6560] \\
J1735$^{\rm *}$ & 8075  &     6.2241       & 0.1612         & -1.76         & -1.2899           &-1.9957 \\
\hline
\end{tabular}
\label{tab:tablafinal}

{\footnotesize
Note: $^{\rm *}$ Solution with canonical H envelope.}
\end{table*}

\begin{table*}[t]
\centering
\caption{Comparison between the main characteristics of the adopted 
asteroseismological models for every ELMV WD shown in this work with the results obtained in \cite{2017A&A...607A..33C}.}
\begin{tabular}{cccccccc}
\hline
\hline
Star & Work &$T_{\rm eff}$ [K]  &$\log(g)$ [cgs]  &$M_{\star}[M_{\sun}]$ &$\log(M_{\rm H}/M_{\star})$ &$\log(R_{\star}/R_{\sun})$ &$\log(L_{\star}/L_{\sun})$ \\
\hline
J1112   &This    & 9301           &     5.9695       & 0.1612          &-1.76                    &-1.1623                   &-1.4932\\
     & Previous  & 9300           &     6.9215       & 0.2390          &-2.45                    &-1.5528                  &-2.2757 \\
\hline           
J1518  &This     & 9487           &     6.9994       & 0.2390          &-3.67                    &-1.5916                  &-2.3200 \\
     & Previous  & 9789           &     7.0956       & 0.2707          &-2.96                    & -1.6126                 &-2.3098 \\
\hline
J1738 & This   &[8883,9273]      &[6.0506,6.6923] &[0.1612,0.1921]  &[-5.43,-1.76]          &[-1.5057,-1.2029]       &[-2.2548,-1.6560] \\
     & Previous  & 9177           &     7.6241       & 0.4352          &-3.21                    & -1.7746    &  -2.7447 \\
\hline
\end{tabular}
\label{tab:comparison}
\end{table*}

\begin{acknowledgements}
We wish to thank our anonymous referee for the constructive
comments and suggestions that greatly improved the original version of
the paper. 
Part of this work was supported by AGENCIA through the Programa de
Modernizaci\'on Tecnol\'ogica BID 1728/OC-AR, and by the PIP
112-200801-00940 grant from CONICET. This research has made use of
NASA Astrophysics Data System.    
\end{acknowledgements}


\bibliographystyle{aa} 
\bibliography{paper-elm-pul-vi} 

\begin{thebibliography}{77}
\expandafter\ifx\csname natexlab\endcsname\relax\def\natexlab#1{#1}\fi

\bibitem[{{Althaus} \& {C{\'o}rsico}(2004)}]{2004A&A...417.1115A}
{Althaus}, L.~G. \& {C{\'o}rsico}, A.~H. 2004, \aap, 417, 1115

\bibitem[{{Althaus} {et~al.}(2010){Althaus}, {C{\'o}rsico}, {Isern}, \&
  {Garc{\'{\i}}a-Berro}}]{2010A&ARv..18..471A}
{Althaus}, L.~G., {C{\'o}rsico}, A.~H., {Isern}, J., \& {Garc{\'{\i}}a-Berro},
  E. 2010, \aapr, 18, 471

\bibitem[{{Althaus} {et~al.}(2013){Althaus}, {Miller Bertolami}, \&
  {C{\'o}rsico}}]{2013A&A...557A..19A}
{Althaus}, L.~G., {Miller Bertolami}, M.~M., \& {C{\'o}rsico}, A.~H. 2013,
  \aap, 557, A19

\bibitem[{{Bell} {et~al.}(2017){Bell}, {Gianninas}, {Hermes}, {Winget},
  {Kilic}, {Montgomery}, {Castanheira}, {Vanderbosch}, {Winget}, \&
  {Brown}}]{2017ApJ...835..180B}
{Bell}, K.~J., {Gianninas}, A., {Hermes}, J.~J., {et~al.} 2017, \apj, 835, 180

\bibitem[{{Bell} {et~al.}(2015){Bell}, {Kepler}, {Montgomery}, {Hermes},
  {Harrold}, \& {Winget}}]{2015ASPC..493..217B}
{Bell}, K.~J., {Kepler}, S.~O., {Montgomery}, M.~H., {et~al.} 2015, in
  Astronomical Society of the Pacific Conference Series, Vol. 493, 19th
  European Workshop on White Dwarfs, ed. P.~{Dufour}, P.~{Bergeron}, \&
  G.~{Fontaine}, 217

\bibitem[{{Bell} {et~al.}(2018){Bell}, {Pelisoli}, {Kepler}, {Brown}, {Winget},
  {Winget}, {Vanderbosch}, {Castanheira}, {Hermes}, {Montgomery}, \&
  {Koester}}]{2018arXiv180511129B}
{Bell}, K.~J., {Pelisoli}, I., {Kepler}, S.~O., {et~al.} 2018, ArXiv e-prints

\bibitem[{{Bogn{\'a}r} {et~al.}(2014){Bogn{\'a}r}, {Papar{\'o}}, {C{\'o}rsico},
  {Kepler}, \& {Gy{\H o}rffy}}]{2014A&A...570A.116B}
{Bogn{\'a}r}, Z., {Papar{\'o}}, M., {C{\'o}rsico}, A.~H., {Kepler}, S.~O., \&
  {Gy{\H o}rffy}, {\'A}. 2014, \aap, 570, A116

\bibitem[{{Bogn{\'a}r} {et~al.}(2016){Bogn{\'a}r}, {Papar{\'o}}, {Moln{\'a}r},
  {P{\'a}pics}, {Plachy}, {Vereb{\'e}lyi}, \&
  {S{\'o}dor}}]{2016MNRAS.461.4059B}
{Bogn{\'a}r}, Z., {Papar{\'o}}, M., {Moln{\'a}r}, L., {et~al.} 2016, \mnras,
  461, 4059

\bibitem[{{Bradley}(1998)}]{1998ApJS..116..307B}
{Bradley}, P.~A. 1998, \apjs, 116, 307

\bibitem[{{Bradley}(2001)}]{2001ApJ...552..326B}
---. 2001, \apj, 552, 326

\bibitem[{{Bradley} {et~al.}(1993){Bradley}, {Winget}, \&
  {Wood}}]{1993ApJ...406..661B}
{Bradley}, P.~A., {Winget}, D.~E., \& {Wood}, M.~A. 1993, \apj, 406, 661

\bibitem[{{Brassard} {et~al.}(1992{\natexlab{a}}){Brassard}, {Fontaine},
  {Wesemael}, \& {Hansen}}]{1992ApJS...80..369B}
{Brassard}, P., {Fontaine}, G., {Wesemael}, F., \& {Hansen}, C.~J.
  1992{\natexlab{a}}, \apjs, 80, 369

\bibitem[{{Brassard} {et~al.}(1991){Brassard}, {Fontaine}, {Wesemael},
  {Kawaler}, \& {Tassoul}}]{1991ApJ...367..601B}
{Brassard}, P., {Fontaine}, G., {Wesemael}, F., {Kawaler}, S.~D., \& {Tassoul},
  M. 1991, \apj, 367, 601

\bibitem[{{Brassard} {et~al.}(1992{\natexlab{b}}){Brassard}, {Fontaine},
  {Wesemael}, \& {Tassoul}}]{1992ApJS...81..747B}
{Brassard}, P., {Fontaine}, G., {Wesemael}, F., \& {Tassoul}, M.
  1992{\natexlab{b}}, \apjs, 81, 747

\bibitem[{{Brown} {et~al.}(2016){Brown}, {Gianninas}, {Kilic}, {Kenyon}, \&
  {Allende Prieto}}]{2016ApJ...818..155B}
{Brown}, W.~R., {Gianninas}, A., {Kilic}, M., {Kenyon}, S.~J., \& {Allende
  Prieto}, C. 2016, \apj, 818, 155

\bibitem[{{Brown} {et~al.}(2010){Brown}, {Kilic}, {Allende Prieto}, \&
  {Kenyon}}]{2010ApJ...723.1072B}
{Brown}, W.~R., {Kilic}, M., {Allende Prieto}, C., \& {Kenyon}, S.~J. 2010,
  \apj, 723, 1072

\bibitem[{{Brown} {et~al.}(2012){Brown}, {Kilic}, {Allende Prieto}, \&
  {Kenyon}}]{2012ApJ...744..142B}
---. 2012, \apj, 744, 142

\bibitem[{{Brown} {et~al.}(2017){Brown}, {Kilic}, {Kosakowski}, \&
  {Gianninas}}]{2017ApJ...847...10B}
{Brown}, W.~R., {Kilic}, M., {Kosakowski}, A., \& {Gianninas}, A. 2017, \apj,
  847, 10

\bibitem[{{Burgers}(1969)}]{1969fecg.book.....B}
{Burgers}, J.~M. 1969, {Flow Equations for Composite Gases} (New York: Academic
  Press)

\bibitem[{{Calcaferro} {et~al.}(2018){Calcaferro}, {Althaus}, \&
  {C{\'o}rsico}}]{2018A&A...614A..49C}
{Calcaferro}, L.~M., {Althaus}, L.~G., \& {C{\'o}rsico}, A.~H. 2018, \aap, 614,
  A49

\bibitem[{{Calcaferro} {et~al.}(2016){Calcaferro}, {C{\'o}rsico}, \&
  {Althaus}}]{2016A&A...589A..40C}
{Calcaferro}, L.~M., {C{\'o}rsico}, A.~H., \& {Althaus}, L.~G. 2016, \aap, 589,
  A40

\bibitem[{{Calcaferro} {et~al.}(2017{\natexlab{a}}){Calcaferro}, {C{\'o}rsico},
  \& {Althaus}}]{2017A&A...600A..73C}
---. 2017{\natexlab{a}}, \aap, 600, A73

\bibitem[{{Calcaferro} {et~al.}(2017{\natexlab{b}}){Calcaferro}, {C{\'o}rsico},
  \& {Althaus}}]{2017A&A...607A..33C}
---. 2017{\natexlab{b}}, \aap, 607, A33

\bibitem[{{Clayton} {et~al.}(2017){Clayton}, {Podsiadlowski}, {Ivanova}, \&
  {Justham}}]{2017MNRAS.470.1788C}
{Clayton}, M., {Podsiadlowski}, P., {Ivanova}, N., \& {Justham}, S. 2017,
  \mnras, 470, 1788

\bibitem[{{C{\'o}rsico} \& {Althaus}(2006)}]{2006A&A...454..863C}
{C{\'o}rsico}, A.~H. \& {Althaus}, L.~G. 2006, \aap, 454, 863

\bibitem[{{C{\'o}rsico} \& {Althaus}(2014)}]{2014A&A...569A.106C}
---. 2014, \aap, 569, A106

\bibitem[{{C{\'o}rsico} \& {Althaus}(2016)}]{2016A&A...585A...1C}
---. 2016, \aap, 585, A1

\bibitem[{{C{\'o}rsico} {et~al.}(2002){C{\'o}rsico}, {Althaus}, {Benvenuto}, \&
  {Serenelli}}]{2002A&A...387..531C}
{C{\'o}rsico}, A.~H., {Althaus}, L.~G., {Benvenuto}, O.~G., \& {Serenelli},
  A.~M. 2002, \aap, 387, 531

\bibitem[{{C{\'o}rsico} {et~al.}(2008){C{\'o}rsico}, {Althaus}, {Kepler},
  {Costa}, \& {Miller Bertolami}}]{2008A&A...478..869C}
{C{\'o}rsico}, A.~H., {Althaus}, L.~G., {Kepler}, S.~O., {Costa}, J.~E.~S., \&
  {Miller Bertolami}, M.~M. 2008, \aap, 478, 869

\bibitem[{{C{\'o}rsico} {et~al.}(2006){C{\'o}rsico}, {Althaus}, \& {Miller
  Bertolami}}]{2006A&A...458..259C}
{C{\'o}rsico}, A.~H., {Althaus}, L.~G., \& {Miller Bertolami}, M.~M. 2006,
  \aap, 458, 259

\bibitem[{{C{\'o}rsico} {et~al.}(2012{\natexlab{a}}){C{\'o}rsico}, {Althaus},
  {Miller Bertolami}, \& {Bischoff-Kim}}]{2012A&A...541A..42C}
{C{\'o}rsico}, A.~H., {Althaus}, L.~G., {Miller Bertolami}, M.~M., \&
  {Bischoff-Kim}, A. 2012{\natexlab{a}}, \aap, 541, A42

\bibitem[{{C{\'o}rsico} {et~al.}(2009){C{\'o}rsico}, {Althaus}, {Miller
  Bertolami}, \& {Garc{\'{\i}}a-Berro}}]{2009A&A...499..257C}
{C{\'o}rsico}, A.~H., {Althaus}, L.~G., {Miller Bertolami}, M.~M., \&
  {Garc{\'{\i}}a-Berro}, E. 2009, \aap, 499, 257

\bibitem[{{C{\'o}rsico} {et~al.}(2014){C{\'o}rsico}, {Althaus}, {Miller
  Bertolami}, {Kepler}, \& {Garc{\'{\i}}a-Berro}}]{2014JCAP...08..054C}
{C{\'o}rsico}, A.~H., {Althaus}, L.~G., {Miller Bertolami}, M.~M., {Kepler},
  S.~O., \& {Garc{\'{\i}}a-Berro}, E. 2014, \jcap, 8, 054

\bibitem[{{C{\'o}rsico} {et~al.}(2007{\natexlab{a}}){C{\'o}rsico}, {Althaus},
  {Miller Bertolami}, \& {Werner}}]{2007A&A...461.1095C}
{C{\'o}rsico}, A.~H., {Althaus}, L.~G., {Miller Bertolami}, M.~M., \& {Werner},
  K. 2007{\natexlab{a}}, \aap, 461, 1095

\bibitem[{{C{\'o}rsico} {et~al.}(2016){C{\'o}rsico}, {Althaus}, {Serenelli},
  {Kepler}, {Jeffery}, \& {Corti}}]{2016A&A...588A..74C}
{C{\'o}rsico}, A.~H., {Althaus}, L.~G., {Serenelli}, A.~M., {et~al.} 2016,
  \aap, 588, A74

\bibitem[{{C{\'o}rsico} {et~al.}(2007{\natexlab{b}}){C{\'o}rsico}, {Miller
  Bertolami}, {Althaus}, {Vauclair}, \& {Werner}}]{2007A&A...475..619C}
{C{\'o}rsico}, A.~H., {Miller Bertolami}, M.~M., {Althaus}, L.~G., {Vauclair},
  G., \& {Werner}, K. 2007{\natexlab{b}}, \aap, 475, 619

\bibitem[{{C{\'o}rsico} {et~al.}(2012{\natexlab{b}}){C{\'o}rsico}, {Romero},
  {Althaus}, \& {Hermes}}]{2012A&A...547A..96C}
{C{\'o}rsico}, A.~H., {Romero}, A.~D., {Althaus}, L.~G., \& {Hermes}, J.~J.
  2012{\natexlab{b}}, \aap, 547, A96

\bibitem[{{De Ger{\'o}nimo} {et~al.}(2017){De Ger{\'o}nimo}, {Althaus},
  {C{\'o}rsico}, {Romero}, \& {Kepler}}]{2017A&A...599A..21D}
{De Ger{\'o}nimo}, F.~C., {Althaus}, L.~G., {C{\'o}rsico}, A.~H., {Romero},
  A.~D., \& {Kepler}, S.~O. 2017, \aap, 599, A21

\bibitem[{{De Ger{\'o}nimo} {et~al.}(2018){De Ger{\'o}nimo}, {Althaus},
  {C{\'o}rsico}, {Romero}, \& {Kepler}}]{2018A&A...613A..46D}
---. 2018, \aap, 613, A46

\bibitem[{{De Ger{\'o}nimo} {et~al.}(2015){De Ger{\'o}nimo}, {C{\'o}rsico},
  {Althaus}, \& {Romero}}]{2015ASPC..493..225D}
{De Ger{\'o}nimo}, F.~C., {C{\'o}rsico}, A.~H., {Althaus}, L.~G., \& {Romero},
  A.~D. 2015, in Astronomical Society of the Pacific Conference Series, Vol.
  493, 19th European Workshop on White Dwarfs, ed. P.~{Dufour}, P.~{Bergeron},
  \& G.~{Fontaine}, 225

\bibitem[{{De Marco} \& {Soker}(2002)}]{2002PASP..114..602D}
{De Marco}, O. \& {Soker}, N. 2002, \pasp, 114, 602

\bibitem[{{Fontaine} \& {Brassard}(2008)}]{2008PASP..120.1043F}
{Fontaine}, G. \& {Brassard}, P. 2008, \pasp, 120, 1043

\bibitem[{{Giammichele} {et~al.}(2017{\natexlab{a}}){Giammichele}, {Charpinet},
  {Brassard}, \& {Fontaine}}]{2017A&A...598A.109G}
{Giammichele}, N., {Charpinet}, S., {Brassard}, P., \& {Fontaine}, G.
  2017{\natexlab{a}}, \aap, 598, A109

\bibitem[{{Giammichele} {et~al.}(2017{\natexlab{b}}){Giammichele}, {Charpinet},
  {Fontaine}, \& {Brassard}}]{2017ApJ...834..136G}
{Giammichele}, N., {Charpinet}, S., {Fontaine}, G., \& {Brassard}, P.
  2017{\natexlab{b}}, \apj, 834, 136

\bibitem[{{Giammichele} {et~al.}(2018){Giammichele}, {Charpinet}, {Fontaine},
  {Brassard}, {Green}, {Van Grootel}, {Bergeron}, {Zong}, \&
  {Dupret}}]{2018Natur.554...73G}
{Giammichele}, N., {Charpinet}, S., {Fontaine}, G., {et~al.} 2018, \nat, 554,
  73

\bibitem[{{Giammichele} {et~al.}(2016){Giammichele}, {Fontaine}, {Brassard}, \&
  {Charpinet}}]{2016ApJS..223...10G}
{Giammichele}, N., {Fontaine}, G., {Brassard}, P., \& {Charpinet}, S. 2016,
  \apjs, 223, 10

\bibitem[{{Gianninas} {et~al.}(2015){Gianninas}, {Kilic}, {Brown}, {Canton}, \&
  {Kenyon}}]{2015ApJ...812..167G}
{Gianninas}, A., {Kilic}, M., {Brown}, W.~R., {Canton}, P., \& {Kenyon}, S.~J.
  2015, \apj, 812, 167

\bibitem[{{Hermes} {et~al.}(2013{\natexlab{a}}){Hermes}, {Montgomery},
  {Gianninas}, {Winget}, {Brown}, {Harrold}, {Bell}, {Kenyon}, {Kilic}, \&
  {Castanheira}}]{2013MNRAS.436.3573H}
{Hermes}, J.~J., {Montgomery}, M.~H., {Gianninas}, A., {et~al.}
  2013{\natexlab{a}}, \mnras, 436, 3573

\bibitem[{{Hermes} {et~al.}(2013{\natexlab{b}}){Hermes}, {Montgomery},
  {Winget}, {Brown}, {Gianninas}, {Kilic}, {Kenyon}, {Bell}, \&
  {Harrold}}]{2013ApJ...765..102H}
{Hermes}, J.~J., {Montgomery}, M.~H., {Winget}, D.~E., {et~al.}
  2013{\natexlab{b}}, \apj, 765, 102

\bibitem[{{Hermes} {et~al.}(2012){Hermes}, {Montgomery}, {Winget}, {Brown},
  {Kilic}, \& {Kenyon}}]{2012ApJ...750L..28H}
---. 2012, \apjl, 750, L28

\bibitem[{{Istrate} {et~al.}(2016){Istrate}, {Marchant}, {Tauris}, {Langer},
  {Stancliffe}, \& {Grassitelli}}]{2016A&A...595A..35I}
{Istrate}, A.~G., {Marchant}, P., {Tauris}, T.~M., {et~al.} 2016, \aap, 595,
  A35

\bibitem[{{Ivanova} \& {Nandez}(2016)}]{2016MNRAS.462..362I}
{Ivanova}, N. \& {Nandez}, J.~L.~A. 2016, \mnras, 462, 362

\bibitem[{{Kepler} {et~al.}(2014){Kepler}, {Fraga}, {Winget}, {Bell},
  {C{\'o}rsico}, \& {Werner}}]{2014MNRAS.442.2278K}
{Kepler}, S.~O., {Fraga}, L., {Winget}, D.~E., {et~al.} 2014, \mnras, 442, 2278

\bibitem[{{Kepler} {et~al.}(2016){Kepler}, {Pelisoli}, {Koester}, {Ourique},
  {Romero}, {Reindl}, {Kleinman}, {Eisenstein}, {Valois}, \&
  {Amaral}}]{2016MNRAS.455.3413K}
{Kepler}, S.~O., {Pelisoli}, I., {Koester}, D., {et~al.} 2016, \mnras, 455,
  3413

\bibitem[{{Kepler} {et~al.}(2012){Kepler}, {Pelisoli}, {Pe{\c c}anha}, {Costa},
  {Fraga}, {Hermes}, {Winget}, {Castanheira}, {C{\'o}rsico}, {Romero},
  {Althaus}, {Kleinman}, {Nitta}, {Koester}, {K{\"u}lebi}, {Jordan}, \&
  {Kanaan}}]{2012ApJ...757..177K}
{Kepler}, S.~O., {Pelisoli}, I., {Pe{\c c}anha}, V., {et~al.} 2012, \apj, 757,
  177

\bibitem[{{Kepler} \& {Romero}(2017)}]{2017EPJWC.15201011K}
{Kepler}, S.~O. \& {Romero}, A.~D. 2017, in European Physical Journal Web of
  Conferences, Vol. 152, European Physical Journal Web of Conferences, 01011

\bibitem[{{Kilic} {et~al.}(2011){Kilic}, {Brown}, {Allende Prieto},
  {Ag{\"u}eros}, {Heinke}, \& {Kenyon}}]{2011ApJ...727....3K}
{Kilic}, M., {Brown}, W.~R., {Allende Prieto}, C., {et~al.} 2011, \apj, 727, 3

\bibitem[{{Kilic} {et~al.}(2012){Kilic}, {Brown}, {Allende Prieto}, {Kenyon},
  {Heinke}, {Ag{\"u}eros}, \& {Kleinman}}]{2012ApJ...751..141K}
---. 2012, \apj, 751, 141

\bibitem[{{Kilic} {et~al.}(2018){Kilic}, {Hermes}, {Corsico}, {Kosakowski},
  {Brown}, {Antoniadis}, {Calcaferro}, {Gianninas}, {Althaus}, \&
  {Green}}]{2018arXiv180603650K}
{Kilic}, M., {Hermes}, J.~J., {Corsico}, A.~H., {et~al.} 2018, ArXiv e-prints

\bibitem[{{Kilic} {et~al.}(2015){Kilic}, {Hermes}, {Gianninas}, \&
  {Brown}}]{2015MNRAS.446L..26K}
{Kilic}, M., {Hermes}, J.~J., {Gianninas}, A., \& {Brown}, W.~R. 2015, \mnras,
  446, L26

\bibitem[{{Kleinman} {et~al.}(1998){Kleinman}, {Nather}, {Winget}, {Clemens},
  {Bradley}, {Kanaan}, {Provencal}, {Claver}, {Watson}, {Yanagida}, {Nitta},
  {Dixson}, {Wood}, {Grauer}, {Hine}, {Fontaine}, {Liebert}, {Sullivan},
  {Wickramasinghe}, {Achilleos}, {Marar}, {Seetha}, {Ashoka}, {Mei{\v s}tas},
  {Leibowitz}, {Moskalik}, {Krzesi{\'n}ski}, {Solheim}, {Bruvold},
  {O'Donoghue}, {Kurtz}, {Warner}, {Martinez}, {Vauclair}, {Dolez},
  {Chevreton}, {Barstow}, {Kepler}, {Giovannini}, {Augusteijn}, {Hansen}, \&
  {Kawaler}}]{1998ApJ...495..424K}
{Kleinman}, S.~J., {Nather}, R.~E., {Winget}, D.~E., {et~al.} 1998, \apj, 495,
  424

\bibitem[{{Koester} {et~al.}(2009){Koester}, {Voss}, {Napiwotzki},
  {Christlieb}, {Homeier}, {Lisker}, {Reimers}, \&
  {Heber}}]{2009A&A...505..441K}
{Koester}, D., {Voss}, B., {Napiwotzki}, R., {et~al.} 2009, \aap, 505, 441

\bibitem[{{Nandez} \& {Ivanova}(2016)}]{2016MNRAS.460.3992N}
{Nandez}, J.~L.~A. \& {Ivanova}, N. 2016, \mnras, 460, 3992

\bibitem[{{Nelemans} \& {Tauris}(1998)}]{1998A&A...335L..85N}
{Nelemans}, G. \& {Tauris}, T.~M. 1998, \aap, 335, L85

\bibitem[{{Papar{\'o}} {et~al.}(2013){Papar{\'o}}, {Bogn{\'a}r}, {Plachy},
  {Moln{\'a}r}, \& {Bradley}}]{2013MNRAS.432..598P}
{Papar{\'o}}, M., {Bogn{\'a}r}, Z., {Plachy}, E., {Moln{\'a}r}, L., \&
  {Bradley}, P.~A. 2013, \mnras, 432, 598

\bibitem[{{Pech} \& {Vauclair}(2006)}]{2006A&A...453..219P}
{Pech}, D. \& {Vauclair}, G. 2006, \aap, 453, 219

\bibitem[{{Pech} {et~al.}(2006){Pech}, {Vauclair}, \&
  {Dolez}}]{2006A&A...446..223P}
{Pech}, D., {Vauclair}, G., \& {Dolez}, N. 2006, \aap, 446, 223

\bibitem[{{Pelisoli} {et~al.}(2018{\natexlab{a}}){Pelisoli}, {Bell}, {Kepler},
  \& {Koester}}]{2018arXiv180504070P}
{Pelisoli}, I., {Bell}, K.~J., {Kepler}, S.~O., \& {Koester}, D.
  2018{\natexlab{a}}, ArXiv e-prints

\bibitem[{{Pelisoli} {et~al.}(2018{\natexlab{b}}){Pelisoli}, {Kepler},
  {Koester}, {Castanheira}, {Romero}, \& {Fraga}}]{2018MNRAS.478..867P}
{Pelisoli}, I., {Kepler}, S.~O., {Koester}, D., {et~al.} 2018{\natexlab{b}},
  \mnras, 478, 867

\bibitem[{{Romero} {et~al.}(2012){Romero}, {C{\'o}rsico}, {Althaus}, {Kepler},
  {Castanheira}, \& {Miller Bertolami}}]{2012MNRAS.420.1462R}
{Romero}, A.~D., {C{\'o}rsico}, A.~H., {Althaus}, L.~G., {et~al.} 2012, \mnras,
  420, 1462

\bibitem[{{Romero} {et~al.}(2017){Romero}, {C{\'o}rsico}, {Castanheira}, {De
  Ger{\'o}nimo}, {Kepler}, {Koester}, {Kawka}, {Althaus}, {Hermes}, {Bonato},
  \& {Gianninas}}]{2017ApJ...851...60R}
{Romero}, A.~D., {C{\'o}rsico}, A.~H., {Castanheira}, B.~G., {et~al.} 2017,
  \apj, 851, 60

\bibitem[{{Romero} {et~al.}(2013){Romero}, {Kepler}, {C{\'o}rsico}, {Althaus},
  \& {Fraga}}]{2013ApJ...779...58R}
{Romero}, A.~D., {Kepler}, S.~O., {C{\'o}rsico}, A.~H., {Althaus}, L.~G., \&
  {Fraga}, L. 2013, \apj, 779, 58

\bibitem[{{Sabach} \& {Soker}(2017)}]{2017arXiv170608897S}
{Sabach}, E. \& {Soker}, N. 2017, ArXiv e-prints

\bibitem[{{Steinfadt} {et~al.}(2010){Steinfadt}, {Bildsten}, \&
  {Arras}}]{2010ApJ...718..441S}
{Steinfadt}, J.~D.~R., {Bildsten}, L., \& {Arras}, P. 2010, \apj, 718, 441

\bibitem[{{Tassoul} {et~al.}(1990){Tassoul}, {Fontaine}, \&
  {Winget}}]{1990ApJS...72..335T}
{Tassoul}, M., {Fontaine}, G., \& {Winget}, D.~E. 1990, \apjs, 72, 335

\bibitem[{{Tremblay} {et~al.}(2015){Tremblay}, {Gianninas}, {Kilic}, {Ludwig},
  {Steffen}, {Freytag}, \& {Hermes}}]{2015ApJ...809..148T}
{Tremblay}, P.-E., {Gianninas}, A., {Kilic}, M., {et~al.} 2015, \apj, 809, 148

\bibitem[{{Winget} \& {Kepler}(2008)}]{2008ARA&A..46..157W}
{Winget}, D.~E. \& {Kepler}, S.~O. 2008, \araa, 46, 157

\end{thebibliography}

\end{document}